\documentclass[journal=jpclcd, manuscript=article, layout=twocolumn]{achemso}
\usepackage[font=small]{caption}  
\usepackage{amsmath}
\usepackage{amssymb}
\usepackage{xcolor}

\usepackage{varioref}		
\usepackage{hyperref}
	\hypersetup{colorlinks=true, urlcolor=black, linkcolor=black, citecolor=black} 			

\newcommand{\visc}{\eta} 

\newcommand{\fks}{\lambda} 

\newcommand{\fkz}{\ell} 

\newcommand{\packdens}{\phi}

\newcommand{\SIsecSimDetails}{S1} 
\newcommand{\SIsecPosDepFric}{S2} 
\newcommand{\SIsecNavierStokes}{S3} 
\newcommand{\SIsecPosDepVisc}{S4} 
\newcommand{\SIsecVeloTailFits}{S5}
\newcommand{\SIsecProfsCalc}{S6}
\newcommand{\SIsecNumStokes}{S7} 
\newcommand{\SIsecVeloFitsLinResp}{S8}
\newcommand{\SIsecVeloReconst}{S9}
\newcommand{\SIsecNaiveVeloModels}{S10}
\newcommand{\SIsecCA}{S11} 
\newcommand{\SIsecEffVisc}{S12} 
\newcommand{\SIsecGKVisc}{S13} 
\newcommand{\SIsecHB}{S14} 
\newcommand{\SIsecGDS}{S15} 
\newcommand{\SIsecDepLen}{S16} 
\newcommand{\SIsecFricWettingTheo}{S17} 

\newcommand{\eq}{eq} 
\newcommand{\eqs}{eqs} 
\newcommand{\Eq}{Equation} 
\newcommand{\Eqs}{Equations} 

\newcommand{\fig}{Figure} 
\newcommand{\figs}{Figures} 

\newcommand{\eqr}{\ref} 
\newcommand{\refcite}[1]{\citenum{#1}}

\title{Sub-Nanometer Interfacial Hydrodynamics: \\The Interplay of Interfacial Viscosity and Surface Friction}
\author{Shane R.\ Carlson}
\author{Roland R.\ Netz}

\affiliation{Fachbereich Physik, Freie Universit{\"a}t Berlin, Arnimallee 14, 14195 Berlin, Germany}
\email{rnetz@physik.fu-berlin.de}
\date{August 12, 2025}


\begin{document}

\maketitle

\begin{abstract}
For an accurate description of nanofluidic systems, it is crucial to account for the transport properties of liquids at surfaces on sub-nanometer scales, where classical hydrodynamics fails due to the finite range of surface--liquid interactions and modifications of the local viscosity.
We show how to account for both via generalized, position-dependent surface-friction and interfacial viscosity profiles, which enables the accurate description of interfacial flow on the nanoscale using the Stokes equation.
Such profiles are extracted from non-equilibrium molecular dynamics simulations of water on polar, non-polar, fluorinated, and unfluorinated alkane and alcohol self-assembled monolayers of widely varying wetting characteristics.
Power-law relationships among the Navier friction coefficient, interfacial viscosity excess, and depletion length are revealed, and these are each found to be exponential in the work of adhesion.
Our framework forms the basis for describing sub-nanometer fluid flow at interfaces with implications for electrokinetics, biophysics, and nanofluidics.

\begin{tocentry}
    \centering
    \includegraphics{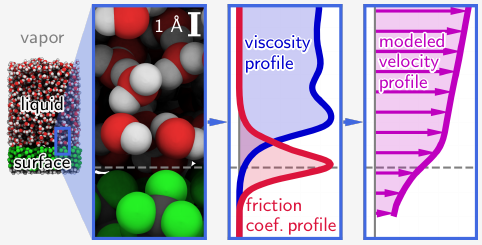}
\end{tocentry}
\end{abstract}

\medskip

The nanoscale flow of liquids at surfaces, especially water, is of crucial importance in electrokinetics \cite{2006_Kim_JCP, 2006_vanderHeyden_NL, 2017_Uematsu_CPL, 2024_Becker_CR}, as well as cell biology and biophysics \cite{2008_Ball_CR, 2018_Huber_CR}, 
such as for transport in transmembrane pores \cite{2019_Trofimov_IJMS}, 
bacterial motility \cite{2016_Lauga_ARFM},
and hydration layers around biomolecules \cite{2017_Laage_CR}.
It is the cornerstone of nanofluidics, which is of great interest recently \cite{2021_Kavokine_ARFM},
with direct applications including carbon nanotube technologies \cite{2001_Hummer_Nature, 2005_Majumder_Nat, 2006_Holt_Sci, 2016_Secchi_Nature}, 
fabricated nanopores and nanoslits \cite{2017_Esfandiar_Science, 2025_Zhang_CRPS}, 
clean energy \cite{2014_Park_CSR, 2021_Zhang_NRM}, 
nanofiltration \cite{2017_Wang_ACIE, 2021_Peng_AFM}, 
and biological/medical applications \cite{2014_Segerink_LC}, 
such as the manipulation of DNA \cite{2013_He_M}, 
intracellular delivery of biomolecules \cite{2021_Hur_AS},
and the analysis of cells, vesicles, and viruses \cite{2021_Yamamoto_AC}.
We seek to understand, down to the sub-nanometer scale, how liquids flow adjacent to surfaces, for which the relevant transport properties are the viscosity and surface--liquid friction.
As one moves to the sub-nanometer scale, surface-to-volume ratios grow very large, which means that interfacial properties (e.g., surface--liquid friction), become as important as bulk properties (e.g., viscosity) \cite{2007_Bocquet_SM, 2025_Shuvo_Nanoscale}.
The surface--liquid friction is characterized for steady-state flows by the Navier friction coefficient, $\fks$, defined via
\begin{equation}
	F_f = -\fks u_{\rm slip}\,,
	\label{eq:slip_coeff_defn}
\end{equation}
where $F_f$ is the surface--liquid friction stress, and $u_{\rm slip}$ the slip velocity of the liquid, i.e., the relative velocity of the liquid directly adjacent to the surface \cite{incoll_1823_Navier}.
Alternatively, the slip may be cast in terms of the slip length $b$, which satisfies $\fks = \visc / b$,
where $\visc$ is the shear viscosity of the liquid \cite{incoll_1860_Helmholtz}. 
In the macroscopic context, the slip length is useful for describing low-friction interfaces, such as gases
\cite{1879_Maxwell_PTRSL}, or very-high-viscosity liquids, such as polymer melts\cite{1993_Migler_PRL}, at solid surfaces.
For liquids like water, slip lengths are typically on the order of nanometers, and the effect of slippage on flow increases with $b/h$, the ratio of the slip length to the channel size \cite{2007_Bocquet_SM}, so slip lengths for macroscopic flows are vanishingly insignificant, and a no-slip condition is sufficient \cite{2005_Neto_RPP}.
Despite this, small slippage at solid--liquid interfaces was measured as early as the 1950s in glass capillary channels \cite{1956_Schnell_JAP, 1959_Debye_JAP}. 
More recently, liquids have been shown to flow through carbon nanotubes at rates multiple orders of magnitude larger than would a no-slip flow \cite{2001_Hummer_Nature, 2005_Majumder_Nat, 2006_Holt_Sci, 2016_Secchi_Nature}.
Note that \eq~\eqr{eq:slip_coeff_defn}  describes the linear-response regime, i.e.\ where the slip velocity is low; when it becomes large enough, the relationship is nonlinear \cite{1997_Thompson_Nature, 2008_Martini_PRL}.

The Navier friction coefficient $\fks$ and/or slip length $b$ can be extracted from non-equilibrium molecular dynamics (NEMD) simulations where shear is induced by applying external driving forces \cite{1997_Thompson_Nature, 1989_Koplik_PFA, 1999_Barrat_PRL, 2001_Cieplak_PRL, 2023_Lafon_PRE}. 
A typical approach is to induce a steady-state Couette or Poiseuille flow and linearly extrapolate the tangential liquid velocity profile $u(z)$ past the interfacial position $z_{\rm int}$ to find the slip length $b$ via
\begin{equation}
	u_{\rm slip} = u(z_{\rm int}) = b \left. \frac{du(z)}{dz} \right\rvert_{z=z_{\rm int}}	\,,
	\label{eq:slip_length_defn_geom} 
\end{equation}
with $z_{\rm int}$ defined for example, as the (fixed) position of the solid surface \cite{1999_Barrat_PRL} or as the Gibbs dividing surface of the liquid \cite{2017_Schlaich_NL,2023_Lafon_PRE}.
While this method works reasonably well for weak surface--liquid interactions and low-friction surfaces,
problems arise for high-friction interfaces where $u_{\rm slip}$ and $b$ are small, difficult to measure, and sensitively dependent on the definition of the interfacial position, $z_{\rm int}$.

In another approach, equilibrium molecular dynamics (MD) simulations of a solid surface with an adsorbed liquid are carried out, the friction coefficient of the center-of-mass coordinate of the liquid or surface are extracted using Green--Kubo relations, and the Navier friction coefficient is found using Stokes' equation, where a position- and time-independent viscosity $\visc$ is assumed \cite{2021_Oga_PRR, 2023_Oga_JCP}.

These models crucially ignore the non-locality of intermolecular interactions, which leads to two modifications of classical hydrodynamics:
firstly, surface--liquid friction acts over a finite distance and is therefore delocalized along $z$, as opposed to acting just at a specific position $z_{\rm int}$,
and secondly, the structure and behavior of liquid near an interface is modified, which means that shear viscosity near interfaces is in general position dependent.
Indeed, viscosity near interfaces has been found to be enhanced when liquid interactions with the adjacent phase are strong, and reduced when they are weak \cite{2009_Sendner_Langmuir, 2017_Schlaich_NL, 2023_Malgaretti_JCP, 2024_Jedlovszky_JCP}. 
Modified interfacial viscosity is of particular importance for nanoconfined liquids \cite{2004_Tas_APL, 2007_Goertz_Langmuir}.
Thus, to fully describe interfacial hydrodynamics at the nanoscale, we take into account the position dependence of the surface--liquid friction coefficient $\fkz(z)$ and interfacial viscosity $\visc(z)$, extracting both from simulations.

\begin{figure*}[!t]
	\centering
	\includegraphics{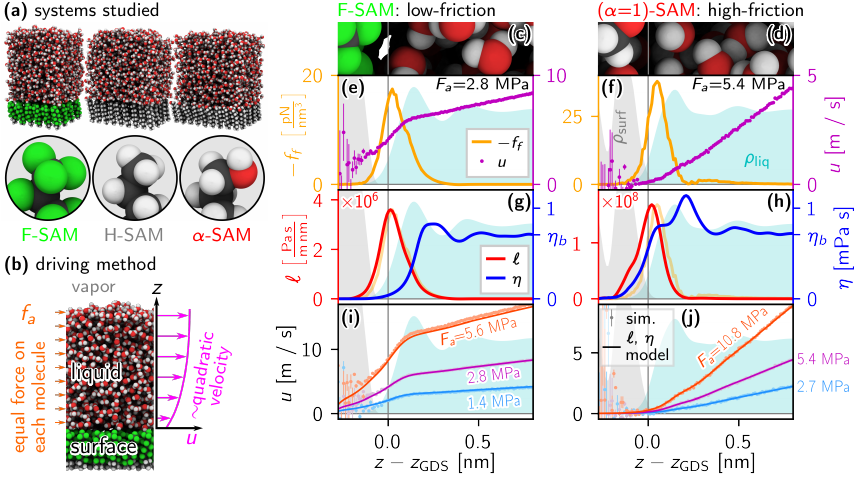}
	\caption{
	\textbf{(a)} Snapshots of the SAM surfaces studied, along with close-ups of the SAM molecule head groups. The SAMs comprise either decane with the top eight carbons perfluorinated (F-SAM), plain decane (H-SAM), or decanol with the charges of the terminal OH-group varied ($\alpha$-SAMs). 
	\textbf{(b)} Schematic illustrating the NEMD method used. The SAM is restrained at the bottom, while a constant force acts equally on each liquid atom.
	\textbf{(c--j)} Surface-friction and viscosity profile extraction for the low-friction F-SAM/water system (left column) and high-friction ($\alpha$=1)-SAM/water system (right column).
	The mass densities of the surface $\rho_{\rm surf}$ and liquid $\rho_{\rm liq}$ are plotted as shaded areas with arbitrary $y$-scaling in all plots. 
	Plots are over $z$, relative to the water Gibbs dividing surface position $z_{\rm GDS}$.
	\textbf{(c, d)} Simulation snapshots for each respective system.
	\textbf{(e, f)} Plots of extracted surface--liquid friction force, $f_f(z)$, and liquid velocity, $u(z)$, profiles.
	The total applied driving stress $F_a$ (see \eq~\eqr{eq:applied_force_calc}) is given in each plot.
	\textbf{(g, h)} Plots of the surface-friction profiles $\fkz(z)$, and viscosity profiles $\visc(z)$, which are calculated from $f_f(z)$ and $u(z)$ profiles (see \eqs~\eqr{eq:ff_tdep_ss} and \eqr{eq:visc_prof_calc}, respectively). 
	The force profiles $f_f(z)$ are also plotted schematically as translucent orange curves for comparison to  $\fkz(z)$.
	The viscosity profiles are calculated from driven-flow simulations with larger driving stresses: $F_a=11.2$~MPa for the F-SAM, and $F_a=21.6$~MPa for the ($\alpha$=1)-SAM.
	\textbf{(i, j)} Velocity profiles $u(z)$ extracted from driven-flow simulations with different applied driving stresses (points with error bars), compared with those calculated using $\fkz(z)$ and $\visc(z)$ from the row above by solving \eq~\eqr{eq:stokes_ss_reconst1} numerically (solid lines).
	\label{fig:workflow_fig}	}
\end{figure*}

To this end, we carry out driven-flow NEMD simulations of liquid water slabs on self-assembled monolayer (SAM) surfaces with widely varying wetting and friction properties.
Above each water slab is vacuum in which a water vapor phase may form.
This open system geometry circumvents the strong sensitivity of viscosity to nanochannel width arising from the incommensurability of discrete liquid-molecule layers---an important phenomenon, but not the focus of this study \cite{2016_NeekAmal_ACSN, 2020_Shadloo_KARJ}.
The SAMs are comprised of decane (H-SAM), decane with the top eight carbons perfluorinated (F-SAM), or decanol ($\alpha$-SAM), where the partial charges on the OH groups are scaled by a factor $\alpha \in \{0, \, 0.5,\, 0.6, \, 0.7, \, 0.8, \, 0.9, \, 1\}$.
The systems are illustrated in \fig~\ref{fig:workflow_fig}(a), though the production simulation systems are larger than those shown. 

The simulations are carried out in GPU-enabled single-precision GROMACS ${2023.3}$ \cite{1995_Berendsen_CPC, 2001_Lindahl_MMA}, using the leap-frog integrator \cite{1974_Hockney_JCP} with a $2$-fs time step.
The velocity-rescaling (CSVR) thermostat is applied to all atoms with a target temperature of $300$~K \cite{2007_Bussi_JCP}.
Because longer force cutoffs are superior for modeling interfacial properties, the Lennard--Jones forces are modeled with force-switching between $1.9$ and $2$~nm \cite{2021_Carlson_Langmuir, 2024_Carlson_JPCL}.
Electrostatic forces are modeled using particle-mesh Ewald beyond a real-space cutoff of $2$~nm \cite{1993_Darden_JCP}.
SAM molecules are modeled using the OPLS All-Atom (OPLS-AA) force field \cite{1984_Jorgensen_JACS, 1994_Kaminski_JPC, 1996_Jorgensen_JACS} with selected  dihedrals optimized \cite{2021_Carlson_Langmuir}. 
Water is modeled using the SPC/E water model \cite{1987_Berendsen_JPC}.
Further simulation details can be found in the SI Section 
{\SIsecSimDetails}.

The water is driven along the $x$-direction, tangential to the SAM, by a constant gravity-like force acting on all liquid atoms, i.e., the force on each atom is proportional to its mass.
In the bulk, where the liquid density is constant, this results in a constant force density, and a quadratic flow profile with the vertex of the parabola at the liquid--vapor interface, i.e., a half-Poiseuille flow, as illustrated in \fig~\ref{fig:workflow_fig}(b).

\figs~\ref{fig:workflow_fig}(c--j) illustrate the workflow for extraction of surface-friction and interfacial-viscosity profiles from the driven-flow simulations, and resulting flows for the low-friction, fluorinated F-SAM (left column) and the high-friction, polar ($\alpha$=1)-SAM (right column). 
\figs~\ref{fig:workflow_fig}(c, d) show snapshots of the simulated systems, zoomed in on the respective interfacial regions of interest.
\figs~\ref{fig:workflow_fig}(e, f) plot the extracted profiles $u(z)$, the liquid velocity, and $f_f(z)$, the friction force per unit volume on the centers of mass of liquid molecules due just to the surface (and \emph{not} to adjacent liquid), as a function of the $z$-position of liquid molecule centers of mass.
The slip at the low-friction, hydrophobic F-SAM is apparent in $u(z)$, i.e., $u(z)$ is clearly positive valued even approaching the surface.
For the $\alpha$-SAM on the other hand, slip is not apparent, and $u(z)$ is convex near the interface, making the definition of a slip velocity or slip length difficult.
The non-locality of $f_f(z)$ is apparent in \figs~\ref{fig:workflow_fig}(e, f), with the surface--liquid friction force acting on the liquid over several Angstroms along $z$ for both surfaces. 
This illustrates why the surface--liquid friction force should not be ignored in viscosity profile calculations.
The total surface--liquid friction stress $F_f$ is given by
\begin{equation}
	F_f = \int \mathrm{d}z\, f_f(z)\,.
	\label{eq:Ff_of_ff_ss}
\end{equation}
In the linear-friction regime, a local, position-dependent friction coefficient $\fkz(z)$ is defined via
\begin{equation}
	f_f(z) = -\fkz(z) u(z)\,.
	\label{eq:ff_tdep_ss}
\end{equation}
Combining \eqs~\eqr{eq:slip_coeff_defn}, \eqr{eq:Ff_of_ff_ss} and \eqr{eq:ff_tdep_ss} yields
\begin{equation}
	\fks u_{\rm slip} = \int \mathrm{d}z\, \fkz(z) u(z) \,.
	\label{eq:totals_equal_z_deps}
\end{equation}
\Eq~\eqr{eq:totals_equal_z_deps}  holds for arbitrary profiles $u(z)$, including constant $u(z)$, with $u = u_{\rm slip}$, from which follows by comparison with \eq~\eqr{eq:slip_coeff_defn} an expression for the Navier friction coefficient based on the microscopically defined friction profile,
\begin{equation}
	\fks = \int \mathrm{d}z\, \fkz(z) \,,
	\label{eq:fks_is_sum_fkzs}
\end{equation}
which, crucially, does not rely on the definition of the interfacial position $z_{\rm int}$.
\Eqs~\eqr{eq:totals_equal_z_deps} and \eqr{eq:fks_is_sum_fkzs} can be combined to give the corresponding slip velocity,
\begin{equation}
	u_{\rm slip} = \frac{\int \mathrm{d}z\, \fkz(z) u(z)}{\int \mathrm{d}z\, \fkz(z)} \,,
	\label{eq:slip_velo_nano_defn}
\end{equation}
which takes the form of a weighted mean over $u(z)$.
Thus, from \eq~\eqr{eq:ff_tdep_ss}, it follows that the Navier friction law, \eq~\eqr{eq:slip_coeff_defn}, holds for $\fks$ and $u_{\rm slip}$ as defined in \eqs~\eqr{eq:fks_is_sum_fkzs} and \eqr{eq:slip_velo_nano_defn} (see the SI Section 
{\SIsecPosDepFric}
for a more complete derivation).
Surface-friction profiles $\fkz(z)$, calculated via \eq~\eqr{eq:ff_tdep_ss}, are plotted for the F-SAM/water and ($\alpha$=1)-SAM/water systems in \figs~\ref{fig:workflow_fig}(g, h). 
In order that the reader is able to discern the small differences in shape and position between $f_f(z)$ and $\fkz(z)$, the forces $f_f(z)$ are also reproduced in \figs~\ref{fig:workflow_fig}(g, h) as translucent orange curves with an arbitrary scaling factor.
As one might expect, the friction profile for the very hydrophilic ($\alpha$=1)-SAM is (two) orders of magnitude higher than that of the hydrophobic F-SAM.

As mentioned above, viscosity is modified at interfaces.
Let $f_a(z)$ denote the applied driving force density acting on the liquid in the $x$-direction,
\begin{equation}
	f_a(z) = \frac{F_a \rho_{\rm liq}(z) }{\int \mathrm{d}z^\prime \, \rho_{\rm liq}(z^\prime)} \,,
	\label{eq:applied_force_calc}
\end{equation}
where  $\rho_{\rm liq}(z)$ is the liquid density profile and $F_a$ is the total applied driving stress.
Then, the total external force density  on the liquid, $f(z)$, can be decomposed as
\begin{equation}
	f(z) = f_a(z) + f_f(z) \,.
	\label{eq:force_on_liq_decomp}
\end{equation}
Water on the nanoscopic scale has a low Reynolds number, so its motion is described well by the linear Stokes equation, \cite{incoll_1823_Navier, 1845_Stokes_TCPS, 1851_Stokes_TCPS}
\begin{equation}
	f(z)
	=
	- \partial_z \visc(z) \partial_z u(z) \,, 
	\label{eq:stokes_ss}
\end{equation}
where  a local, position-dependent shear viscosity $\visc(z)$ is assumed (see the SI Section
{\SIsecNavierStokes}
for derivations of the Navier--Stokes and Stokes equations for position-dependent viscosity).
Letting $z=z_0$ denote a position below the liquid phase, integration of \eq~\eqr{eq:stokes_ss} from $z_0$ to an arbitrary $z$ gives
\begin{align}
	\visc(z) =
	-\frac{1}{\partial_{z}u(z)}
		\int_{z_0}^{z}\mathrm{d}z^\prime \, f(z^\prime) \,,
		\label{eq:visc_prof_calc}
\end{align}
a more thorough derivation and discussion of which can be found in the SI Section 
{\SIsecPosDepVisc}. 
Viscosity profiles $\visc(z)$, calculated from \eq~\eqr{eq:visc_prof_calc}, are plotted for the F-SAM/water and $\alpha$-SAM/water systems in \figs~\ref{fig:workflow_fig}(g, h) (see the SI Sections 
{\SIsecVeloTailFits} and
{\SIsecProfsCalc}
for details and plots on extraction of $\fkz(z)$ and $\visc(z)$).
Each viscosity profile converges to a bulk value $\eta_b$, which agree well between the two systems.

\begin{figure*}[!h]
	\centering
	\includegraphics{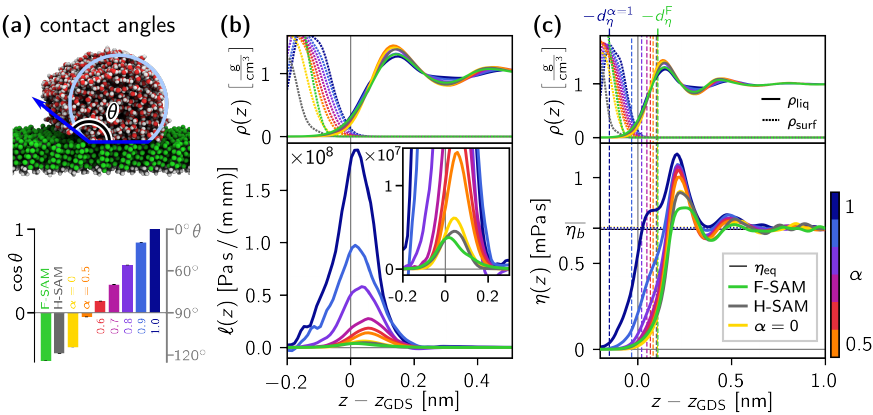}
	\caption{Comparison of surface-friction profiles and viscosity profiles of water adsorbed on different SAM surfaces, plotted relative to the water Gibbs dividing surface position $z_{\rm GDS}$.
	As a positional reference, the mass densities of the surface $\rho_{\rm surf}(z)$ and liquid $\rho_{\rm liq}(z)$ for all systems are plotted in the upper panels as dotted and solid lines, respectively.
	\textbf{(a)} Water contact angles $\theta$, from droplet simulations for all surfaces. Also shown is a schematic illustrating the contact angle $\theta$, overlaid on a simulation snapshot of a cylindrical water droplet on an F-SAM.
	 The ($\alpha$=1)-SAM  was fully wetted by water, so we show it as having a contact angle of $0^\circ$.
	\textbf{(b)}	 Surface-friction profiles $\fkz(z)$ for all systems, obtained via \eq~\eqr{eq:ff_tdep_ss}.
	The inset shows the same data over a smaller range of $y$-values, so that the profiles for the low-friction surfaces are discernible.
	\textbf{(c)} Viscosity profiles $\eta(z)$ obtained via  \eq~\eqr{eq:visc_prof_calc}.
	The average of each viscosity profile in the bulk domain, $\visc_b$ is shown as a horizontal dotted line; these agree well among systems, with an average of $\overline{\visc_b} = 0.706$ mPa$\,$s, and with the viscosity extracted from an equilibrium bulk-water simulation, $\visc_{\rm eq}=0.698$ mPa$\,$s, which is shown as a solid black line.
	The viscosity dividing surfaces $z_\visc$ from each viscosity profile are shown as vertical dashed lines.
	\label{fig:allprofs_fig}	}
\end{figure*}

Several previous publications take a similar approach to \eq~\eqr{eq:visc_prof_calc}, but take $f(z)\propto\rho_{\rm liq}(z)$, which, crucially, does not account for the surface--liquid friction stress \cite{1997_Akhmatskaya_JCP, 2012_Parez_PCCP}. 
Others take very different  approaches, including measuring liquid shear between two oppositely moving surfaces \cite{2009_Sendner_Langmuir, 2013_Bonthuis_JPCB, 2017_Schlaich_NL} or in flow driven by electric fields \cite{2021_WoldeKidan_Langmuir}, via the local pressure tensor \cite{1995_Todd_PRE, 2010_Sofos_IJHMT, 2017_Papanikolaou_PRE}, or via the shear and momentum flux \cite{2009_Muller_CPC}, none of which, however, correctly separate surface--liquid stress and viscous effects.
Ref.~\refcite{2012_Hoang_PRE}, on the other hand, calculates interfacial viscosity profiles for Lennard--Jones fluids via pairwise fluid-fluid interactions, which circumvents surface friction contributions.

Substituting \eqs~\eqr{eq:ff_tdep_ss} and \eqr{eq:force_on_liq_decomp} into the Stokes equation,  \eq~\eqr{eq:stokes_ss}, gives
\begin{equation}
	\fkz(z) u(z) = f_a(z) + \partial_z \visc(z) \partial_{z}u(z) 
	\,.
	\label{eq:stokes_ss_reconst1}
\end{equation}
Having extracted $\fkz(z)$ and $\visc(z)$ for a specific surface--liquid pair at a given temperature and pressure, $u(z)$ may be found by solving \eq~\eqr{eq:stokes_ss_reconst1} numerically for an arbitrarily chosen applied force profile $f_a(z)$ (see the SI Section 
{\SIsecNumStokes}). 
\figs~\ref{fig:workflow_fig}(i, j) compare velocity profiles extracted from driven-flow simulations with those calculated by solving \eq~\eqr{eq:stokes_ss_reconst1}.
For the liquid--vapor boundary, it is assumed that $\partial_{z}u(z)=0$, while at the surface, it is assumed that $u(z)=0$.
The overall modeling of the flow is very accurate, which validates our model, including the locality of $\fkz(z)$ and $\visc(z)$, and indicates that the system is in the linear-response regime for both the viscosity and friction (linear-response is also verified in the SI Section
{\SIsecVeloFitsLinResp}).
The same comparison between the  $\fkz$--$\visc$ model and driven-flow simulation data is carried out for all systems studied in this work in the SI Section
{\SIsecVeloReconst}.
In the SI Section
{\SIsecNaiveVeloModels},
we show that failing to account for the position dependence of either $\fkz(z)$ or $\visc(z)$ leads to models that fail to accurately capture the behavior near the interface. 

The surface friction and viscosity profiles for all systems are collected in \fig~\ref{fig:allprofs_fig}.
Wetting is a natural way to classify surface--liquid interactions, so we extract water contact angles of all systems from density profiles of equilibrium MD simulations of cylindrical droplets\cite{2024_Carlson_JPCL}, and extrapolate the droplet size to the macroscopic limit\cite{2008_Sedlmeier_Biointerphases, 2018_Kanduc_PRE, 2017_Kanduc_JCP_Going, 2021_Carlson_Langmuir}. 
The contact angle extraction is detailed in the SI Section
{\SIsecCA}. 
\fig~\ref{fig:allprofs_fig}(a) shows a simulation snapshot of such a cylindrical water droplet on an F-SAM, with the contact angle indicated schematically, and the resulting contact angles for all systems, which range from the very hydrophobic F-SAM ($\theta = 125^\circ$) to the fully wetted ($\alpha$=1)-SAM.

\fig~\ref{fig:allprofs_fig}(b) shows the surface-friction profiles $\fkz(z)$, with the corresponding mass density profiles shown above as a positional reference.
As the friction varies by more than an order of magnitude, the curves for the low-friction surfaces are difficult to discern in the main plot, and the same data is plotted over a smaller range of $y$-values in the inset.
The surface-friction  profiles are delocalized over several Angstroms, with FWHM ranging from 1.1 to 1.8 {\AA} across all systems.
This reveals that, on sub-nanometer scales, friction should not be treated as acting only at a single position $z_{\rm int}$.

\begin{figure*}[h]
	\centering
	\includegraphics{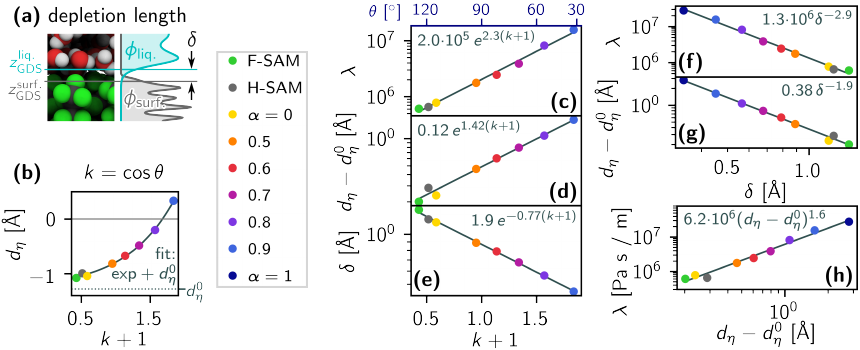}
	\caption{Analysis of relationships among surface--liquid friction, viscosity excess, depletion, and wetting for water adsorbed on different SAM surfaces.
	\textbf{(a)} Schematic illustrating the depletion length, which is the distance between the Gibbs dividing surfaces of the liquid and solid, calculated using the packing density $\packdens$ rather than the mass density $\rho$.
	\textbf{(b)} Viscosity excess distance, $d_\eta$, as extracted in \fig~\ref{fig:allprofs_fig}(c), as a function of $k+1$, where $k=\cos\theta$ is the wetting coefficient. 
	The data are fit with an exponential function plus a constant $d_\eta^0$, which is subtracted from $d_\visc$ in subsequent plots. 
	Negative $d_\visc$ values indicate a viscosity deficit.
	\textbf{(c--e)} Quantities plotted over $k+1$, along with exponential fits of the data. The data for the ($\alpha$=1)-SAM is omitted here, as $k$ is undefined for full wetting.
	\textbf{(c)} The friction coefficient $\lambda$ (in units of Pa$\,$s$\,$m$^{-1}$ throughout the figure), extracted by integrating over each $\fkz(z)$ in \fig~\ref{fig:allprofs_fig}(b). 
	\textbf{(d)} The viscosity excess distance $d_\eta$, less the constant $d_\eta^0$ (see (b)).
	The  $d_\eta-d_\eta^0$ data and fit from (b) are replotted in log-linear to demonstrate that the data fall on a straight line and agree well with the fit.
	\textbf{(e)} The depletion length $\delta$, as illustrated in (b).
	\textbf{(f--h)} Relationships among  $\lambda$, $d_\eta$, and $\delta$ plotted in log-log, alongside power law fits.
	\label{fig:analysis_fig}	}
\end{figure*}

The viscosity profiles $\visc(z)$ are shown in \fig~\ref{fig:allprofs_fig}(c).
Note that these all decay to zero in the region of the SAM, which has been observed before for Lennard--Jones fluids \cite{2012_Hoang_PRE}.
If surface friction is ignored in \eq~\eqr{eq:visc_prof_calc}, i.e., all flow behavior is attributed to the position-dependent viscosity alone, the resulting effective viscosity will typically be found to diverge near the interface (see the SI Section
{\SIsecEffVisc}
for more on effective viscosities, including plots for all systems studied in this work).
The viscosity profiles for all systems converge very near the same value in the bulk, as expected.
For each system, the viscosity value in the bulk, $\eta_b$, is calculated as the mean over the profile in a window between 1.5 and 2.5~nm from the top carbon atoms.
These are shown as dotted horizontal lines and they agree well, all falling within the range $\overline{\eta}_b \pm 0.007$ mPa$\,$s, where $\overline{\eta}_b = 0.706$ mPa$\,$s is their mean value, which agrees well with the bulk value of $\eta_{\rm eq} = 0.698$~mPa$\,$s, calculated via the Green--Kubo relation from an equilibrium  MD simulation where the same water force field is used (see the SI Section
{\SIsecGKVisc} 
for details) \cite{1994_Daivis_JCP, 1998_Alfe_PRL, 2009_Gonzalez_JCP, 2020_Schulz_PRF}.

Alongside the viscosity profiles $\visc(z)$ in \fig~\ref{fig:allprofs_fig}(c), the viscosity dividing surfaces $z_\visc$ (where the viscosity excess vanishes) are shown as vertical dashed lines, and differ from the Gibbs dividing surface $z_{\rm GDS}$ by up to about an Angstrom in either direction.
This reveals that  $z_{\rm GDS}$ is not the relevant interface for liquid viscosity and highlights the importance of treating the position-dependence of the interfacial viscosity on sub-nanometer scales.
We define the interfacial viscosity excess distance as $d_\eta = z_{\rm GDS} - z_\visc$.
Thus, a positive $d_\eta$  indicates a viscosity excess at the interface, while  a negative $d_\eta$ indicates a deficit.
The viscosity excess grows significantly as the surfaces become more hydrophilic, with a shoulder forming in the profile at the first hydration layer nearest the surface, which likely results from the conformational rigidity of water hydrogen-bonded to, or otherwise strongly interacting with, the surface, preventing other water molecules from easily flowing past.
One interesting aspect visible in all the systems is the apparent disagreement between the viscosity and density profiles: naively one might expect viscosity to increase with density, but we observe the main peak of each viscosity profile to be shifted bulkward relative to that of the density profile.
This is likely related to water orientation and especially hydrogen bonding near the interface.
Indeed, we show that there is a deficit of water-water hydrogen-bonded molecules near the interface for all surfaces in the SI Section
{\SIsecHB}.

In \fig~\ref{fig:analysis_fig}, we explore the relationships among the Navier friction coefficient $\fks$, the interfacial viscosity excess distance $d_\eta$, the wetting coefficient $k=\cos\theta$, and the depletion length $\delta$.
The depletion length is the distance between the adjacent Gibbs dividing surfaces of the SAM and the water phases, as illustrated in \fig~\ref{fig:analysis_fig}(a), where the Gibbs dividing surface of a phase is the thermodynamically relevant interfacial position (see the SI Section 
{\SIsecGDS}).
For the SAMs, which are chemically inhomogeneous near the interface, the Gibbs dividing surfaces are calculated as the position where the excess of the packing density $\packdens(z)$ vanishes (see the SI Section
{\SIsecDepLen} 
for further discussion).

For partial wetting, the wetting coefficient $k$ is related to the areal work of adhesion of the surface $W$ via the Young-Dupr{\'e} equation,
\begin{equation}
	W = \gamma_{lv} \left( 1 + k \right) \,,
	\label{eq:young_dupre}
\end{equation}
where $\gamma_{lv}$ is the liquid--vapor interfacial tension of the liquid. 
Thus, we plot $k+1$ rather than $\theta$, and because \eq~\eqr{eq:young_dupre} only holds for partial wetting, we do not include the data for the ($\alpha$=1)-SAM in plots involving $k+1$. 

In \fig~\ref{fig:analysis_fig}(b), $d_\eta$ is plotted over $k+1$ alongside a least-squares fit of an exponential function plus a constant, which we call $d_\eta^0$. 
From the fitted function we calculate $d_\eta(k=-1) = -1.25$~{\AA}, i.e., there is an interfacial viscosity deficit of 1.25~{\AA} in the dewetting limit, where $\theta=180^\circ$, which should be characteristic of the water model alone.
\figs~\ref{fig:analysis_fig}(c--e) are log-linear plots of $\fks$, $d_\eta - d_\eta^0$, and $\delta$ over $k+1$, alongside exponential fits. 
The fits in \figs~\ref{fig:analysis_fig}(c, e) are performed in logarithmic space.

From transition state theory \cite{1940_Kramers_P, 2022_Bruenig_PRE}, it can be shown that $\fks \propto e^{\beta U_0}$ for small driving forces,  where $U_0$ is the barrier height of the corrugated energy landscape seen by a liquid molecule at a surface and $\beta = (k_BT)^{-1}$ is the inverse thermal energy.
From there, it can be shown under certain approximations that
\begin{equation}
	\fks \propto e^{A (k+1)}\,,
	\label{eq:fric_propto_exp_wetting}
\end{equation}
where $A$ is a positive constant.
Indeed, we find $\fks$ to be exponential $k+1$ in \fig~\ref{fig:analysis_fig}(c), where
the fit  gives $A\approx 2.3$. 
Previous works have derived other friction/barrier-height relationships, including  $\fks \propto e^{\beta U_0} / U_0 $ and  $\fks \propto 1+ \left( \beta U_0 \right)^2 / 16$ \cite{2009_Huang_PRL, 2009_Sendner_Langmuir, 2012_Erbas_JACS}.
Fits of these functions to our $\fks(k)$ data, a more detailed derivation of \eq~\eqr{eq:fric_propto_exp_wetting}, and related discussion can be found in the SI Section
{\SIsecFricWettingTheo}.

\fig~\ref{fig:analysis_fig}(d) shows the same fit as (b), but shifted by $-d_\visc^0$ and in log-linear, where the data indeed appear to fall on a straight line, indicating exponential behavior.
\fig~\ref{fig:analysis_fig}(e) plots the depletion length, and these data seem to follow an exponential decay. 

\figs~\ref{fig:analysis_fig}(f--h) examine the relationships among $\fks$, $d_\eta - d_\eta^0$, and $\delta$. 
As these all appeared exponential in $k+1$, we expect them to relate to one another via power laws, and accordingly plot them in log-log.
The data are fitted by taking the logarithm of both datasets and carrying out a linear fit.
From \fig~\ref{fig:analysis_fig}(f) we find $\fks \propto \delta^{-2.9}$, from (g), $d_\eta - d_\eta^0  \propto \delta^{-1.9}$, and from (h), $\fks \propto (d_\eta - d_\eta^0)^{1.6}$.

Thus, we have shown that both surface-friction and inhomogeneous interfacial viscosity profiles are important for understanding flow, and that properly decoupling the two allows for the calculation of useful quantities, such as the surface viscosity excess and Navier friction coefficient, without \textit{a priori} recourse to arbitrarily defined interfacial positions.
Going forward, these methods for accurately decoupling surface--liquid friction and interfacial viscosity will lend themselves well to the analysis of flow on sub-nanometer scales for a wide variety of systems, such as systems with liquid mixtures, nanochannels, and charged surfaces and/or electrolyte solutions, which are relevant for electrokinetics.
The constitutive relations we find between friction coefficient $\fks$, viscosity excess $d_\visc$, depletion length $\delta$, and wetting coefficient $k$ allow for the parameter-free modeling of interfacial flow at smooth surfaces of arbitrary polarity.

\begin{acknowledgement}
The authors thank the Deutsche Forschungsgemeinschaft (DFG, German Research Foundation) for funding via Project-ID 387284271 (SFB 1349).
\end{acknowledgement}

\begin{suppinfo}
S1: Simulation Details.
S2: Position-Dependent Surface--Liquid Friction.
S3: Navier--Stokes and Stokes Equations for Position-Dependent Viscosity.
S4: Calculation of Position-Dependent Viscosity.
S5: Fits of Velocity Profile Tails.
S6: Calculation of Friction and Viscosity Profiles for All Systems.
S7: Numerically Solving the Stokes Equation.
S8: Verifying Linear-Response Regime by Velocity Profile Fits.
S9: Modeling Flow for All Systems.
S10: Other Approaches to Modeling Flow.
S11: Contact Angles.
S12: Effective Viscosity Profiles.
S13: Bulk Shear Viscosity from the Green--Kubo Relation.
S14: Hydrogen Bonding Near the Interface.
S15: Gibbs Dividing Surface.
S16: Depletion Length.
S17: The Friction--Wettability Relationship.
\end{suppinfo}

\begingroup
\small
\bibliography{bibliography.bib}
\endgroup

\end{document}



\newcommand{\mainFigWorkFlow}{1} 
\newcommand{\mainFigAllProfs}{2} 
\newcommand{\mainFigAnalysis}{3} 

\newcommand{\mainEqVeloReconst}{12}

\renewcommand{\thefigure}{S\arabic{figure}}
\renewcommand{\thetable}{S\arabic{table}}
\renewcommand{\theequation}{S\arabic{equation}}
\renewcommand{\thesection}{S\arabic{section}}
\renewcommand{\thesubsection}{S\arabic{section}.\arabic{subsection}}

\renewcommand{\Re}{\,\operatorname{Re}\:}
\renewcommand{\Im}{\,\operatorname{Im}\:}
\newcommand{\ft}{\,\mathcal{F}\:}
\newcommand{\ift}{\,\mathcal{F}^{-1}\:}
\newcommand{\fk}{\Lambda} 
\newcommand{\fkr}{L} 
\newcommand{\fks}{\lambda} 

\newcommand{\mk}{\Gamma} 
\newcommand{\mkr}{G}
\newcommand{\mks}{\gamma}

\newcommand{\fkz}{\ell} 
\newcommand{\fkzs}{\zeta} 
\newcommand{\fkzsu}{l} 

\newcommand{\visc}{\eta} 
\newcommand{\packdens}{\phi} 
\newcommand{\effmass}{M} 

\newcommand{\depfigs}{/home/shane/Documents/SAM/analysis/depletion/plots}
\newcommand{\hbfigs}{/home/shane/Documents/SAM/analysis/surfliq_HB}

\newcommand\red[1]{\textcolor{red}{#1}}

\newcommand{\Eq}{Equation} 
\newcommand{\eq}{eq} 
\newcommand{\eqs}{eqs} 

\newcommand{\fig}{Figure} 
\newcommand{\figs}{Figures} 

\newcommand{\eqr}{\ref}



\title{{\bfseries Supporting Information:} Sub-Nanometer Interfacial Hydrodynamics: The Interplay of Interfacial Viscosity and Surface Friction}

\author{Shane R.\ Carlson}
\author{Roland R.\ Netz\thanks{rnetz@physik.fu-berlin.com}}
\affil[1]{Fachbereich Physik, Freie Universit{\"a}t Berlin, Arnimallee 14, 14195 Berlin, Germany}
\date{August 12, 2025}

\maketitle

\section{Simulation Details}

\begin{figure*}[!t]
	\centering
	\includegraphics{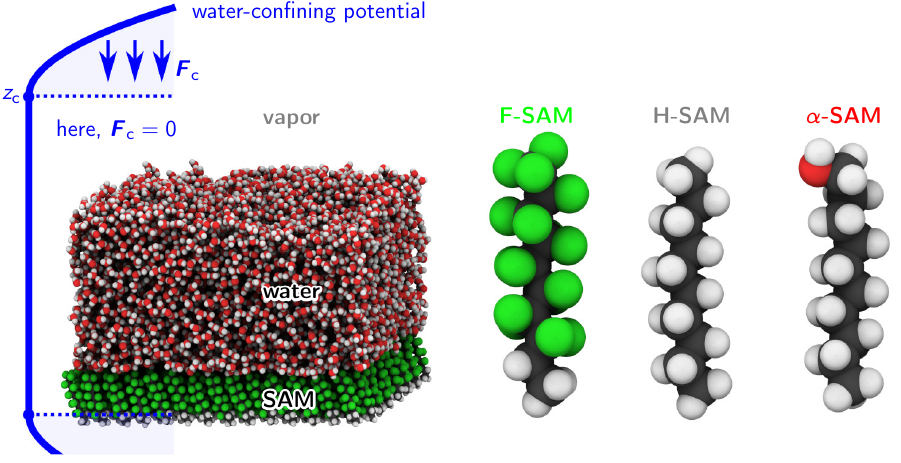}
	\caption{
	\textbf{Left:} A snapshot of the production simulation system for the F-SAM, which consists of a close-packed $14\times 14$ grid of molecules with an 8192-molecule water slab.
	Also shown here schematically is the water-confining potential, which acts with a harmonic force only on water molecules above $z_{\rm c}$.
	\textbf{Right:}
	Snapshots of single SAM molecules for the F-SAM, H-SAM, and an $\alpha$-SAM. 
	Dark gray atoms are C, light gray H, red O, and green F.
	Partial charges are tuned on the OH-groups of the $\alpha$-SAM.
	\label{fig:sys}}
\end{figure*}

We carried out non-equilibrium force-field molecular dynamics (MD) simulations of gravity-driven liquid water slabs on alkane, fluoroalkane, and alcohol self-assembled monolayer (SAM) surfaces.
These surfaces vary widely in their wetting and friction properties.
All system boundaries are periodic, with the surfaces and water slabs contiguous across the boundaries in the $xy$-plane. 
The water slabs consist of 8192 water molecules and are roughly 4~nm thick.
See \fig~\ref{fig:sys} for a snapshot of a simulated system. 
Above each is vacuum in which a water vapor phase may form.

To prevent molecules from crossing the periodic boundary in the $z$-direction and adhering to the bottom of the SAM, a water-confining potential is applied, which is also illustrated in \fig~\ref{fig:sys}.
This potential consists of a flat bottom and harmonic sides and applies only to water molecules and only along the $z$-direction.
The lower harmonic part of the potential is positioned below the SAM, where water molecules do not reach in any case.
The entire liquid water slab sits in the flat bottom of the potential, so the potential does not act on the water molecules there.
The upper harmonic part is positioned above the top of the water slab, and acts on water molecules in the vapor phase that are above position $z_{\rm c}$ with force
\begin{equation}
	\boldsymbol{F}_{\rm c} = -k\left(z - z_{\rm c}\right) \hat{z} \,.
	\label{eq:flatbottom_force}
\end{equation}
This reflects water molecules downward and back toward the water slab.
The harmonic parts have spring constants of \mbox{$k=79.997$ kJ/(mol nm$^2$)} for the water oxygens and \mbox{$k=5.040$ kJ/(mol nm$^2$)} for the water hydrogens, so that an equal acceleration is applied to both atom species.

The SAMs are comprised of decane (H-SAM), decane with the top eight carbons perfluorinated (F-SAM), or decanol ($\alpha$-SAM), where the partial charges on the OH group of the heads are scaled by a factor $\alpha \in  \{0, \, 0.5,\, 0.6, \, 0.7, \, 0.8, \, 0.9, \, 1\}$, which changes the electric dipole.
Snapshots of the molecules are shown in \fig~\ref{fig:sys}.
The bottom carbon atoms of the SAM molecules (the ones furthest from the water slab) are restrained to points in space arranged in a  hexagonal grid in the $xy$-plane via a harmonic potential with a spring constant of \mbox{$k=25000$ kJ/(mol nm$^2$)}.
The nearest-neighbor distance  of the grid points, i.e., the ``grafting distance'', is set to 5.9~{\AA}   for the \mbox{F-SAM} and 4.97~{\AA} for the  H-SAM and $\alpha$-SAMs.
These values are in accordance with experimental findings for F-SAMs and H-SAMs on a gold surface, namely \mbox{Au(111)} \cite{1988_Strong_Langmuir, 1990_Chidsey_Langmuir, 1989_Ulman_Langmuir, 1993_Alves_Langmuir, 1994_Liu_JCP, 1996_Jaschke_JPC}.
These SAM grafting distances are discussed at length in Ref.~\refcite{2021_Carlson_Langmuir}.
The H-SAM and $\alpha$-SAMs are $18\times 16$ molecules in a simulation box of  $8.946\times6.887\times40$~nm, while the F-SAM is $14\times14$ in a simulation box of $8.260\times7.153\times40$~nm. 

The simulations were carried out in GPU-enabled single-precision GROMACS version ${2023.3}$ \cite{1995_Berendsen_CPC, 2001_Lindahl_MMA}, using the leap-frog integrator \cite{1974_Hockney_JCP} with a time step of $2$ fs.
The velocity-rescaling (CSVR) thermostat is applied to all atoms with a target temperature of $300$~K and a coupling time-constant of {\tt tau-t} of 1~ps \cite{2007_Bussi_JCP}.
Longer force cutoffs have been shown to be superior for modeling interfacial properties \cite{2021_Carlson_Langmuir, 2024_Carlson_JPCL}.
Therefore, the Lennard--Jones forces are modeled using a force-switching scheme between $1.9$ and $2$~nm.
Electrostatic forces are modeled using the particle-mesh Ewald (PME) method beyond the real-space cutoff of $2$~nm \cite{1993_Darden_JCP}.

A force and potential correction for the electrostatic PME is used to account for interactions between system replicas along the $z$-direction by setting {\tt ewald-geometry = 3dc}.
SAM molecules are modeled using the OPLS All-Atom (OPLS-AA) force field \cite{1984_Jorgensen_JACS, 1994_Kaminski_JPC, 1996_Jorgensen_JACS} with selected  dihedrals optimized for the F-SAM \cite{2021_Carlson_Langmuir}. 
All covalent bond lengths involving hydrogen atoms are fixed by setting {\tt constraints = hbonds}.
Water is modeled using the SPC/E water model \cite{1987_Berendsen_JPC}.
SPC/E was originally optimized with a 0.9-nm Lennard--Jones cutoff, and, because a longer cutoff is employed in this work, the liquid properties can be expected to differ.
Table \ref{tab:SPCE_properties} compares the properties of SPC/E water at 300~K for these two Lennard--Jones cutoff schemes: the water density and shear viscosity of the bulk liquid are  virtually unchanged, but as the cutoff increases, the surface tension increases significantly, toward the experimental value of 71.99(36) mJ/m$^2$ at \mbox{$25^\circ$~C} \cite{1983_Vargaftik_JPCRD}.
\begin{table}[ht]
    \centering
    \begin{tabular}{p{19mm} p{23mm} p{23mm}}
    \hline
    & 0.9~nm \newline potential-shift & 1.9--2~nm \newline force-switch \\
    \hline 
	$\rho$ [g/cm$^3$] &   $0.99802(1)$  & $0.99803(1)$  \\ 
	& 0.998 \cite{1987_Berendsen_JPC} &  \\
    $\eta_b$ [mPa$\,$s] &  $0.698(9)$   & $0.698(4)$  \\ 
    $\gamma_{lv}$ [mJ/m$^2$] & $54.379(98)$ \cite{2024_Carlson_JPCL}   & $60.260(60)$ \cite{2024_Carlson_JPCL}  \\
    \hline
    \end{tabular}
    \caption{Properties of SPC/E water with two different LJ cutoff schemes: a potential shift with a cutoff of 0.9~nm (as used in the original SPC/E publication, Ref.~\refcite{1987_Berendsen_JPC}), and force switching between 1.9 and 2~nm (as used elsewhere in this work). 
    Where no reference is given, quantities were calculated by us.}
    \label{tab:SPCE_properties}
\end{table}

The water is driven along the $x$-direction, tangential to the SAM by a gravity-like force, where a force is applied to each liquid atom that is proportional to its mass.
In the bulk, the liquid density is constant, which gives a constant force density and a quadratic flow profile with the vertex of the parabola at the liquid--vapor interface, i.e., a half-Poiseuille flow.

The surface--liquid force on each liquid atom is needed in order to extract the surface--liquid friction force as a function of liquid position, $f_f(z)$.
This is extracted by carrying out two reruns with modified topologies. 
In both reruns, all bonded interactions are set to zero. 
In the first, all nonbonded interactions are left on, giving the total nonbonded force on each liquid atom. 
Here, we denote the total nonbonded force on the $i^{\rm th}$ liquid atom as  $F_{i}$.
In the second, all nonbonded interactions involving the surface atoms are set to zero, which gives the nonbonded force on the $i^{\rm th}$ liquid atom from just the other liquid atoms $F_{i}^{\rm liq. \, liq.}$.
Then the force on the $i^{\rm th}$ liquid atom from the surface only is given by 
\begin{equation}
F_{i}^{\rm surf. \, liq.} = F_{i} - F_{i}^{\rm liq. \, liq.} \,.
\end{equation}
Note that the \emph{total} surface--liquid force can be obtained without reruns because
\begin{equation}
	\sum_i F_i^{\rm liq. \, liq.} = 0 \,.
\end{equation}

\section{Position-Dependent Surface--Liquid Friction}
\label{sec:PosDepFric}
Consider a system consisting of a fixed solid surface parallel to the $xy$-plane with an adsorbed liquid phase flowing over the surface along the $x$-direction in a steady-state flow. 
The friction between the surface and liquid is typically characterized by the Navier friction coefficient $\fks$, defined via
\begin{equation}
	F_f = -\fks u_{\rm slip}\,,
	\label{eq:slip_coeff_defn}
\end{equation}
where $F_f$ is the tangential surface--liquid friction stress and $u_{\rm slip}$ the liquid slip velocity, i.e., the velocity of the liquid directly adjacent to the surface, relative to the surface \cite{incoll_1823_Navier}. 
Note that $F_f$ and $u_{\rm slip}$ are both restricted to the $x$-direction here. 
On sub-nanometer scales, $F_f$ acts on liquid over a finite range of $z$-values,
\begin{equation}
	F_f = \int \mathrm{d}z\, f_f(z)\,,
	\label{eq:Ff_of_ff_ss}
\end{equation}
where $f_f(z)$ is the friction force density (along $\hat{x}$) at $z$ due to just the surface (and \emph{not} to adjacent liquid). In the linear-friction regime, a local, steady-state friction coefficient $\fkz(z)$ can be defined,
\begin{equation}
	f_f(z) = -\fkz(z) u(z) \,,
	\label{eq:ff_tdep_ss}
\end{equation}
where $u(z)$ is the velocity of the liquid at $z$. 
Here, $\fkz(z)$ is itself taken to be local, i.e., friction at $z$ is a function only of velocity at $z$ and not at other nearby positions.
This is validated by the accuracy of modeling of interfacial flows for many surface types where a local friction coefficient $\fkz(z)$  is assumed, as shown in \fig~\ref{fig:9velo_reconst}.
Combining \eqs~\eqr{eq:slip_coeff_defn}, \eqr{eq:Ff_of_ff_ss} and \eqr{eq:ff_tdep_ss} yields
\begin{equation}
	\fks u_{\rm slip} = \int \mathrm{d}z\, \fkz(z) u(z) \,.
	\label{eq:totals_equal_z_deps}
\end{equation}
Equation~\eqr{eq:totals_equal_z_deps} holds for arbitrary profiles $u(z)$, including constant $u(z)$, with $u = u_{\rm slip}$, from which follows an expression for the Navier friction coefficient in terms of the microscopically defined friction profile,
\begin{equation}
	\fks = \int \mathrm{d}z\, \fkz(z) \,.
	\label{eq:fks_is_sum_fkzs}
\end{equation}
Together, \eqs~\eqr{eq:totals_equal_z_deps} and \eqr{eq:fks_is_sum_fkzs} give the slip velocity as a weighted mean of $u(z)$,
\begin{equation}
	u_{\rm slip} = \frac{\int \mathrm{d}z\, \fkz(z) u(z)}{\int \mathrm{d}z\, \fkz(z)} \,,
	\label{eq:slip_velo_nano_defn}
\end{equation}
for which \eq~\eqr{eq:slip_coeff_defn} holds, even for position-dependent surface--liquid friction.
For the limiting case of a localized friction coefficient acting at one position $z_{\rm int}$, i.e., \mbox{$\fkz(z) \propto \delta(z-z_{\rm int})$}, \eq~\eqr{eq:fks_is_sum_fkzs} implies that \mbox{$\fkz(z) = \fks \delta(z-z_{\rm int})$}, which together with \eq~\eqr{eq:slip_velo_nano_defn} in turn implies that $u_{\rm slip}=u(z_{\rm int})$. 
Thus, we recover \eq~\eqr{eq:slip_coeff_defn} under the classical assumptions of Navier, and \eqs~\eqr{eq:fks_is_sum_fkzs} and \eqr{eq:slip_velo_nano_defn} can be thought of as giving generalizations of the Navier friction coefficient $\fks$ and slip velocity $u_{\rm slip}$, respectively.

Combining \eqs~\eqr{eq:ff_tdep_ss} and \eqr{eq:fks_is_sum_fkzs} yields a useful equation for the Navier friction coefficient in the case of a steady state flow,
\begin{equation}
	\fks = 
	-\int \mathrm{d}z\, \frac{f_f(z)}{u(z)}\,.
	\label{eq:lambda_ss}
\end{equation}
\Eq~\eqr{eq:lambda_ss} allows $\fks$ to be calculated directly from a simulation where a liquid is driven with a constant external force  tangential to a fixed surface. 

\section{Navier--Stokes and Stokes Equations for Position-Dependent Viscosity}
\label{sec:NavierStokes}
The $i^{\mathrm{th}}$ component of the conservation of momentum in an arbitrary volume $V$ may be expressed via the equation
\begin{align}
	&\frac{d}{dt} \int_V \mathrm{d}\boldsymbol{r} \, \rho u_i 
	+ \int_{\partial V} \mathrm{d} S_j \, \rho u_i u_j \notag \\
	&{} \qquad =
	\int_V  \mathrm{d}\boldsymbol{r} \, f_i
	+ \int_{\partial V} \mathrm{d} S_j \, \sigma_{ij}
	\,.
	\label{eq:cons_momentum}
\end{align}
The four terms, starting on the left, give the change in momentum, momentum flux, external force, and internal force on a surface element. 
The Einstein summation convention is used and arguments are omitted from the following functions for brevity: density $\rho(\boldsymbol{r}, t)$, velocity $\boldsymbol{u}(\boldsymbol{r}, t)$, force density $\boldsymbol{f}(\boldsymbol{r}, t)$, and stress tensor $\sigma_{ij}(\boldsymbol{r}, t)$. 
Gauss's law may be used to write the surface integrals in \eqr{eq:cons_momentum} as volume integrals over divergences, yielding
\begin{equation}
	\int_V  \mathrm{d}\boldsymbol{r} \, 
	\left(
		\frac{\partial }{\partial t} \rho u_i
		+ \nabla_j \rho u_i u_j
		- f_i
		- \nabla_j \sigma_{ij}
	\right)
	= 0\,.
	\label{eq:gauss}
\end{equation}
This holds for arbitrarily defined volumes $V$, so
\begin{align}
	f_i	
	+ \nabla_j \sigma_{ij}
	&=
	\frac{\partial }{\partial t} \rho u_i
	+ \nabla_j u_j \rho  u_i
	\label{eq:gauss_row2} \\
	&= \rho \dot{u}_i 
	+ u_i \dot{\rho}
	+ u_i \nabla_j u_j \rho 
	+ u_j \rho \nabla_j u_i \,.
	\label{eq:gauss_row3}
\end{align}
Using conservation of mass,
\begin{equation}
	\dot{\rho} + \nabla_j u_j \rho = 0 \,,
	\label{eq:mass_cons}
\end{equation}
to cancel terms in \eqr{eq:gauss_row3}, it follows that
\begin{align}
	f_i 
	+ \nabla_j \sigma_{ij}
	= 
	\rho \left( \dot{u}_i 
	+ u_j \nabla_j u_i  \right)
	\equiv \rho \frac{D u_i}{D t} \,,
	\label{eq:pre_nav_stokes}
\end{align}
where $D/Dt$ is the material or substantial derivative,
\begin{equation}
	\frac{D}{Dt} 
	\equiv 
	\frac{\partial}{\partial t}
	+
	u_j \nabla_j \,,
	\label{eq:material_deriv_defn}
\end{equation}
which describes the change of a quantity for an observer comoving with the flow $\boldsymbol{u}$.
The fluid stress tensor, $\sigma_{ij}$, can only depend on spatial derivatives of the velocity and not on the velocity directly. 
Further, assuming a Newtonian fluid, $\sigma_{ij}$ must be rotationally invariant to linear order. 
We consider the case where the volume and shear viscosities are assumed to be homogeneous in time, but position-dependent, non-local (i.e., stress at $\boldsymbol{r}$ is also a function of motion at other positions $\boldsymbol{r} - \boldsymbol{r}^\prime$), and having memory (i.e., stress at time $t$ is also a function of motion at earlier times $t-t^\prime$).
In this case, the stress tensor $\sigma_{ij}$ is given by the linear stress constitutive equation 
\begin{align}
	&\sigma_{ij}(\boldsymbol{r}, t)
	=  
	- p(\boldsymbol{r}, t) \delta_{ij} \label{eq:presstens_genl} \\
	&+ \int\mathrm{d}\boldsymbol{r}^\prime \!\! \int_{0}^{\infty} \mathrm{d}t^\prime 
	\Bigg[	
	 \delta_{ij} \xi^\dagger(\boldsymbol{r}, \boldsymbol{r}^\prime, t^\prime) 
	 \nabla_k u_k(\boldsymbol{r} - \boldsymbol{r}^\prime, t - t^\prime) 
	\notag  \\ 
	&+ \eta^\dagger(\boldsymbol{r}, \boldsymbol{r}^\prime, t^\prime)
	\bigg( 
	\nabla_i u_j(\boldsymbol{r} - \boldsymbol{r}^\prime, t - t^\prime) 
	\notag  \\ 
	&\hspace{2.37cm}+ \nabla_j u_i(\boldsymbol{r} - \boldsymbol{r}^\prime, t - t^\prime) 
	\notag  \\ 
	&\hspace{2.37cm}-\frac{2}{3} \delta_{ij} 
	\nabla_k u_k(\boldsymbol{r} - \boldsymbol{r}^\prime, t - t^\prime) 
	\bigg) \Bigg]\,, \notag 
\end{align}
where $p(\boldsymbol{r}, t)$ is the pressure and $\xi^\dagger(\boldsymbol{r}, \boldsymbol{r}^\prime, t^\prime)$ and $\visc^\dagger(\boldsymbol{r}, \boldsymbol{r}^\prime, t^\prime)$ are the volume- and shear-viscosity kernels, respectively \cite{2025_Kiefer_PRE}.
The terms in \eq~\eqr{eq:presstens_genl} including $p$ and $\xi^\dagger$ are isotropic (diagonal), giving the stress from static pressure and isotropic expansion, respectively. The term including $\eta^\dagger$ is deviatoric (traceless), giving the stress due to shear.
A velocity gradient $\nabla_\alpha u_\beta(\boldsymbol{r} - \boldsymbol{r}^\prime, t- t^\prime)$ may be expanded around $\left(\boldsymbol{r}^\prime, t^\prime \right) = (\vec{0}, 0)$, giving
\begin{align}
	\nabla_\alpha u_\beta(\boldsymbol{r} - \boldsymbol{r}^\prime, t - t^\prime) 
	= 
	&\nabla_\alpha u_\beta(\boldsymbol{r}, t) \notag \\
	&-
	\boldsymbol{r}^\prime \nabla_\alpha \nabla_\gamma u_\beta(\boldsymbol{r}, t) \notag \\
	&-
	t^\prime \nabla_\alpha \dot u_\beta(\boldsymbol{r}, t) \notag \\
	&+ \mathcal{O}\left( \left( \boldsymbol{r}^\prime \right)^2, \left( t^\prime \right)^2, \boldsymbol{r}^\prime  t^\prime \right) \,.
\end{align}
Approximating $\nabla_\alpha u_\beta(\boldsymbol{r} - \boldsymbol{r}^\prime, t - t^\prime)$ by just the leading-order term, $\sigma_{ij}$ may be written 
\begin{align}
	&\sigma_{ij}(\boldsymbol{r}, t)
	=  
	- p(\boldsymbol{r}, t) \delta_{ij}
	+ \xi(\boldsymbol{r}) \delta_{ij} \nabla_k u_k(\boldsymbol{r}, t)
	\label{eq:presstens}\\ 
	&+ \eta(\boldsymbol{r}) \left( 
		\nabla_i u_j(\boldsymbol{r}, t)
		+ \nabla_j u_i(\boldsymbol{r}, t)
		-\frac{2}{3} \delta_{ij} \nabla_k u_k(\boldsymbol{r}, t)
	\right)\,, \notag 
\end{align}
where $ \xi(\boldsymbol{r})$ and $\visc(\boldsymbol{r})$ are local and memoryless volume and shear viscosities, given   by
\begin{align}
	&\xi(\boldsymbol{r}) = 
	\int\mathrm{d}\boldsymbol{r}^\prime \!\! \int_{0}^{\infty} \mathrm{d}t^\prime \,
	\xi^\dagger(\boldsymbol{r}, |\boldsymbol{r}^\prime|, t^\prime) \quad \text{and} \notag\\
	&\visc(\boldsymbol{r}) = 
	\int\mathrm{d}\boldsymbol{r}^\prime \!\! \int_{0}^{\infty} \mathrm{d}t^\prime \,
	\visc^\dagger(\boldsymbol{r}, |\boldsymbol{r}^\prime|, t^\prime) \,.
	\label{eq:r_dep_viscs_defn}
\end{align}
Note that although the viscosities $\xi$ and $\visc$ in \eq~\eqref{eq:r_dep_viscs_defn} are, in general, functions of the position vector $\boldsymbol{r}$, for the systems considered in this work, which are translationally invariant in $x$ and $y$, they depend only on the coordinate $z$.
The assumption of the memorylessness of the shear viscosity is valid for this work, because systems are studied only in a steady state.
The locality of the viscosity, on the other hand, is hypothesized \textit{a priori}. 
This hypothesis is well supported by the accuracy of modeling of interfacial flows for many surface types where a local shear viscosity  is assumed, as shown in \fig~\ref{fig:9velo_reconst}.

Calculating  $\nabla_j \sigma_{ij}$ from  \eq~\eqr{eq:presstens} and substituting the result into  \eq~\eqr{eq:pre_nav_stokes} gives a formulation of the Navier--Stokes equation for position-dependent viscosity, \cite{incoll_1823_Navier, 1845_Stokes_TCPS, 1851_Stokes_TCPS}
\begin{align}
	\rho \frac{D u_i}{D t}
	&= \rho \dot{u_i} + \rho u_j \nabla_j u_i \notag \\
	&=
	f_i
	-\nabla_i p 
	+ \nabla_i \left( \xi - \frac{2 \visc}{3} \right)  \nabla_j u_j 
	 \notag \\
	&\qquad + \nabla_j \visc  \nabla_i u_j
	+ \nabla_j \visc  \nabla_j u_i\,, \label{eq:nav_stokes}
\end{align}
where arguments $\boldsymbol{r}$ and $t$ are again omitted for brevity.
Consider the scaling relations 
$\rho u_j \nabla_j u_i \sim \rho u^2 / L$ and 
$\eta \nabla_k \nabla_k u_i \sim \eta u / L^2$ , where $u$ and $L$ are characteristic speed and length, respectively.
The nonlinear term, $\rho u_j \nabla_j u_i$, is negligible if
\begin{equation}
	\frac{\rho u^2 / L}{\eta u / L}
	=
	\frac{\rho u L}{\eta} 
	\equiv
	\mathrm{Re}
	< 1 \,,
	\label{eq:reynolds_defn}
\end{equation} 
i.e., if the Reynolds number, Re, is small.
For water, $\rho \approx 10^3$ kg/m$^3$ and $\eta \approx 10^{-3}$ kg/m$\,$s, so $\mathrm{Re} < 1$ if $uL  < 10^{-6}$ m$^2$/s, which is fulfilled for the nanoscopic systems of interest to us. Omission of the nonlinear term gives the Stokes' equation,
\begin{align}
	\rho \dot{u_i}
	&=
	f_i
	-\nabla_i p 
	+ \nabla_i \left( \xi - \frac{2 \visc}{3} \right)  \nabla_j u_j
	 \notag \\
	&\qquad + \nabla_j \visc  \nabla_i u_j
	+ \nabla_j \visc  \nabla_j u_i\,. \label{eq:stokes0}
\end{align}%
\section{Calculation of Position-Dependent Viscosity}
\label{sec:PosDepVisc}

We begin by referring to \eq~\eqr{eq:stokes0}, which is the Stokes equation for position-dependent shear and bulk viscosities $\visc(\boldsymbol{r})$ and $\xi(\boldsymbol{r})$.
As in Section \ref{sec:NavierStokes}, $\rho$ is the liquid mass density, $u$ is the liquid velocity, $\dot{u}$ is its time derivative, $f$ is the external force density exerted on the liquid, and the Einstein summation convention is used.
Assume a system translationally invariant in the $xy$-plane.
It immediately follows that $\visc(\boldsymbol{r})$ depends only on  $z$, so we denote it simply as $\visc(z)$.
Next it is assumed that the flow is laminar and parallel, and the applied force $f=f_x(z,t)$ is constant over $x$ and $y$ and acts only in the $x$ direction.
Because the Reynolds number is low, it follows that $u= u_x(z,t)$ and $p=p(z,t)$.
These symmetries cause several terms in \eq~\eqr{eq:stokes0} to vanish, leaving
\begin{equation}
	\rho \dot{u}_x(z,t)
	=
	f_x(z,t)
	+  \partial_z \visc(z) \partial_z u_x(z,t) \,. \label{eq:stokes}
\end{equation}
Next, we assume a steady state flow. 
Omitting the subscripts $x$ for clarity gives
\begin{equation}
	f(z)
	=
	- \partial_z \visc(z) \partial_z u(z) \,. \label{eq:stokes_ss}
\end{equation}
Now consider a system consisting of a planar surface in the $xy$-plane with an adsorbed liquid slab. 
Letting $z_0$ denote a position along $z$ below the liquid phase, integration of \eq~\eqr{eq:stokes_ss} from $z_0$ to an arbitrary $z$ gives
\begin{align}
	\int_{z_0}^{z} \mathrm{d}z^\prime \, f(z^\prime)
	&=
	- \int_{z_0}^{z} \mathrm{d}z^\prime \,
	\partial_{z^\prime} \visc(z^\prime) \partial_{z^\prime} u(z^\prime) \notag \\
	&=
	-\visc(z) \partial_z u(z)
	 \,. \label{eq:int_stokes_ss}
\end{align}
Rearranging this equation gives an expression for the viscosity profile
\begin{align}
	\visc(z) =
	-\frac{1}{\partial_{z} u(z)}
		\int_{z_0}^{z}\mathrm{d}z^\prime \, f(z^\prime) \,, \label{eq:visc_prof_calc}
\end{align}
which can be calculated numerically from driven-flow NEMD simulation data. 
Here, $f(z)$ consists of the sum of the applied driving force density $f_a(z)$  and surface-friction force density $f_f(z)$, and $\partial_{z} u(z)$ is the local liquid shear, which can be obtained numerically from the velocity profile using finite differences.

\begin{figure*}[!ht]
	\centering
	\textsf{Larger Driving Force for Viscosity Profile Extraction:}
	\includegraphics{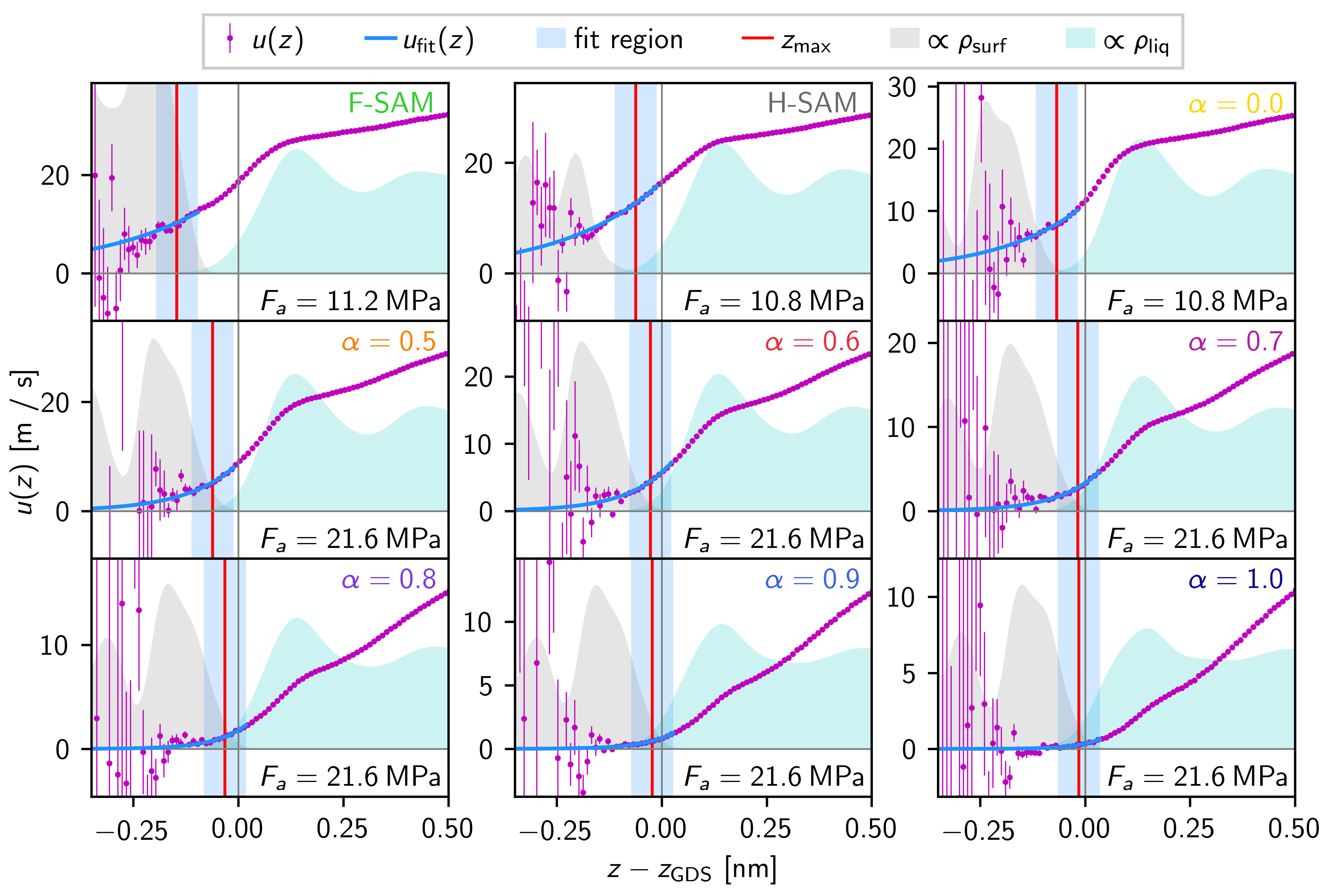}
	\\
	\hspace*{1cm}
	\textsf{Smaller Driving Force for Friction Profile Extraction:}
	\includegraphics{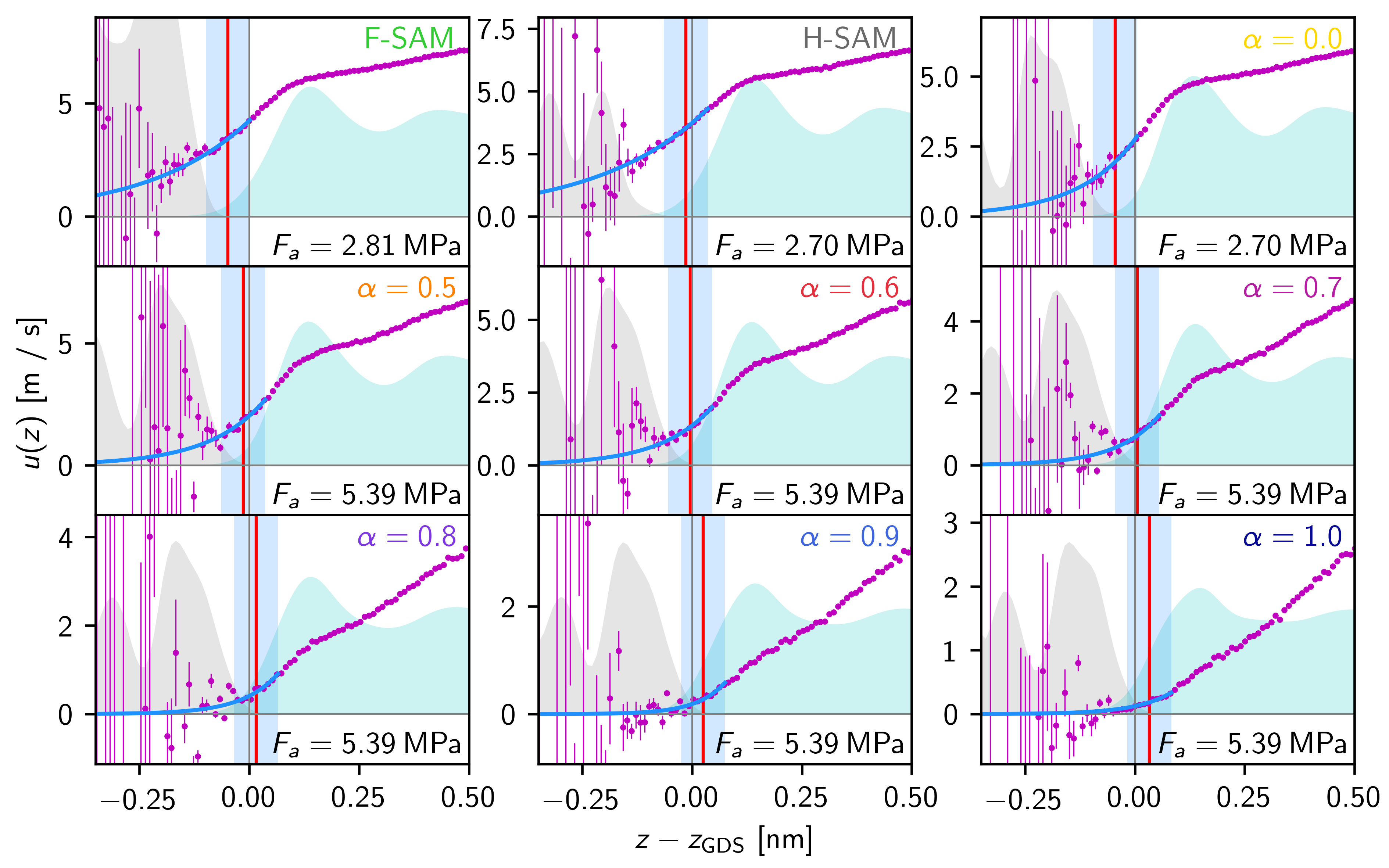}
	\caption{
	Velocity profiles, $u(z)$, shown with exponential fits of decaying tails for two different driving force magnitudes for each surface.
	The upper block of plots shows data from simulations with larger driving forces, which are used to calculate viscosity profiles $\visc(z)$.
	The lower block shows data from simulations with smaller driving forces, which are used to calculate friction coefficient profiles $\fkz(z)$.
	The applied driving stress $F_a$ is printed in each respective panel.
	The water and SAM densities are also shown on an arbitrary scale as shaded cyan and gray areas, respectively.
	The fit region is shown as a blue shaded area and the fitted function is shown as a blue curve.
	All velocity data for $z-z_{\rm GDS} \le z_{\rm max}$, indicated by the vertical red line, is replaced with the fit data before calculating the viscosity or surface liquid friction profiles.
	\label{fig:9_velo_fits}}
\end{figure*}

\section{Fits of Velocity Profile Tails}

In the driven-flow SAM/water simulations, the water velocity tends to decay toward zero as one approaches the SAM surface.
\fig~\ref{fig:9_velo_fits} shows exponential fits of these decaying velocity profile tails.
The upper set of plots are from simulations with a faster driven flow, where the applied driving stress $F_a$ is greater, and are used for extracting viscosity profiles.
The lower set are from simulations with a slower driven flow, i.e., with a smaller applied driving stress $F_a$, and are used for extracting surface--liquid friction coefficients.
Each fit is carried out in just a 1-{\AA} window, which is indicated in the respective plot by a shaded blue area.
The velocity profile to the left of the center of the window (i.e., left of the vertical red line) is replaced by the fit values for the purpose of calculating the viscosity or friction coefficient profiles.
Also shown in each panel are the mass density profiles for the liquid and surface, scaled by an arbitrary factor, as a positional reference.


\begin{figure*}[!t]
	\centering
	\includegraphics{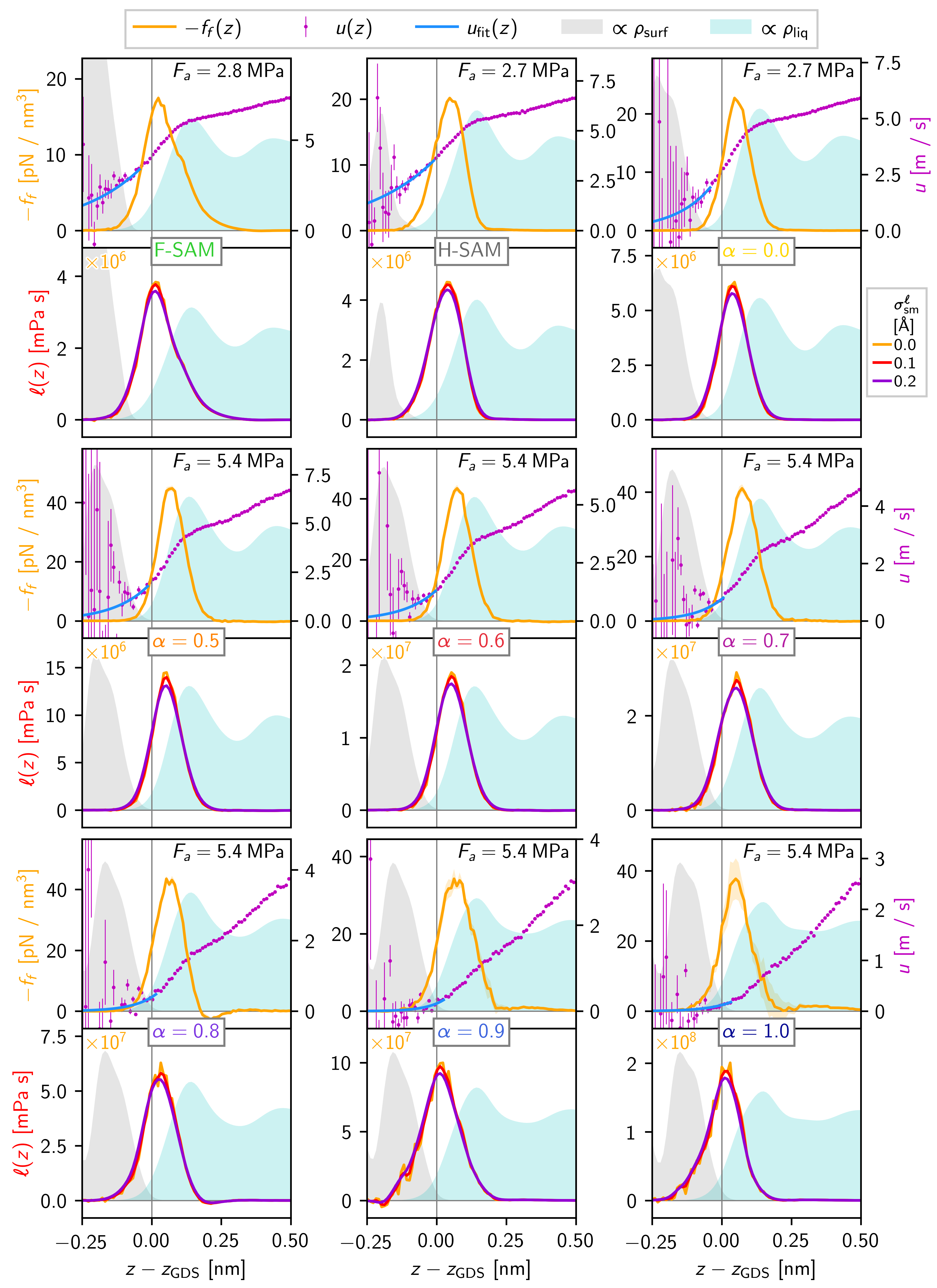}
	\caption{
	Raw velocity $u(z)$ and surface--liquid friction force $f_f(z)$ data (upper panels of each subfigure), and resulting friction coefficient profiles $\fkz(z)$ (lower) for all systems.
	Alongside the velocity profiles $u(z)$, the exponential functions fitted to the tails are shown.
	The friction coefficient profiles $\fkz(z)$  are shown for various degrees of smoothing of the input profiles $u(z)$ and $f_f(z)$.
	The water and SAM densities are also shown on an arbitrary scale as shaded cyan and gray areas, respectively.
	\label{fig:9profiles_gen_fric}}
\end{figure*}

\begin{figure*}[!t]
	\centering
	\includegraphics{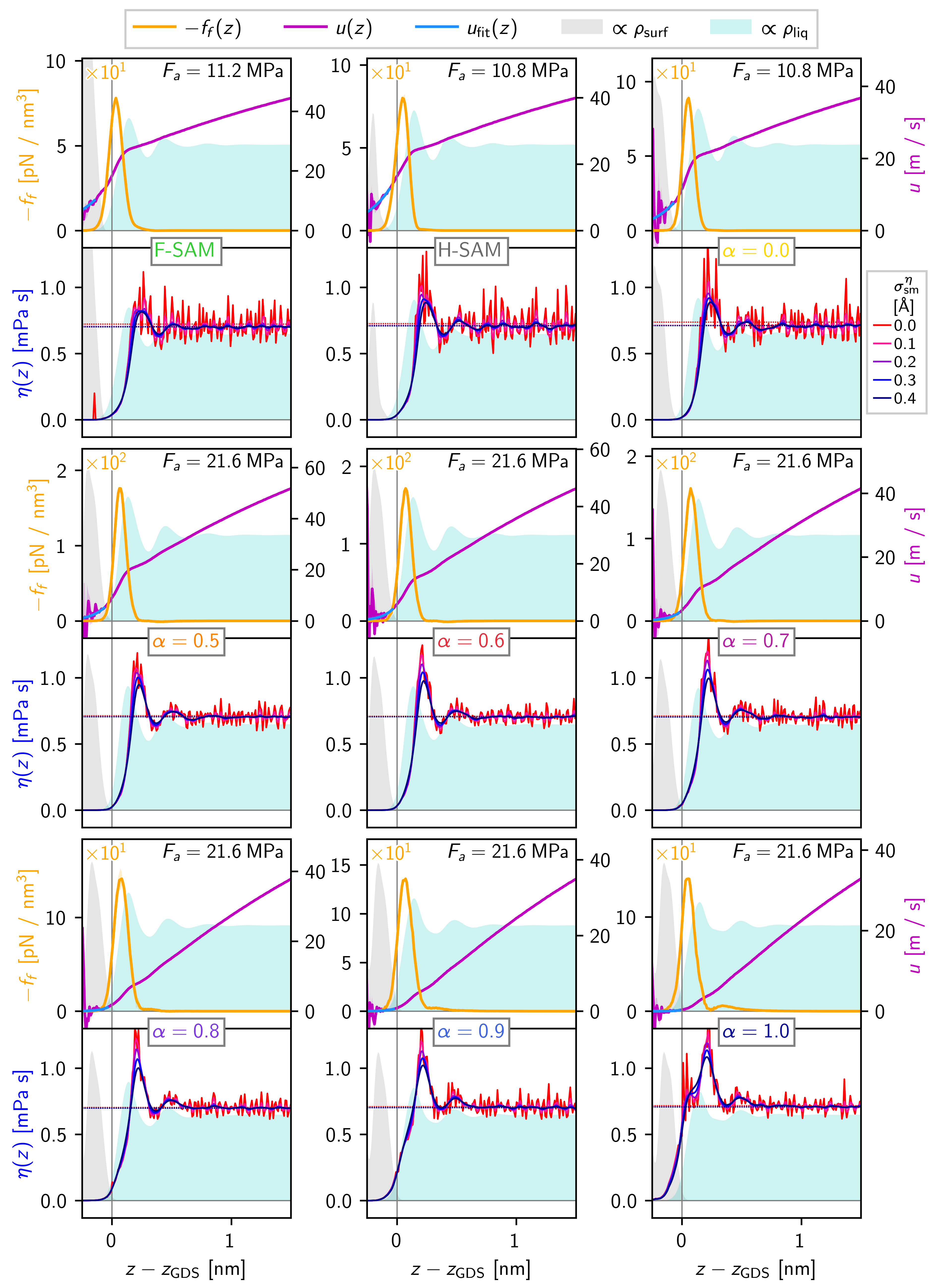}
	\caption{
	Raw velocity $u(z)$ and surface--liquid friction force $f_f(z)$ data (upper panels of each subfigure), and resulting viscosity profiles $\visc(z)$ (lower) for all systems.
	Alongside the velocity profiles $u(z)$, the  exponential functions fitted to the tails are shown.
	The viscosity profiles are shown for various degrees of smoothing of just the input velocity profile $u(z)$.
	The water and SAM densities are also shown on an arbitrary scale as shaded cyan and gray areas, respectively.
	\label{fig:9profiles_gen_visc}}
\end{figure*}

\section{Calculation of Friction and Viscosity Profiles for All \\ Systems}
\label{sec:ProfsCalc}

\fig~\ref{fig:9profiles_gen_fric} shows, for all systems, the raw velocity $u(z)$ and surface--liquid friction force $f_f(z)$ data from driven-flow non-equilibrium molecular dynamics (NEMD) simulations used to extract the surface--liquid friction coefficient profiles $\fkz(z)$, which in turn shown in the lower panels of the subfigures, directly below the corresponding raw data. 
Here, after the fitting procedure illustrated in \fig~\ref{fig:9_velo_fits}, $u(z)$ and $f_f(z)$ are both smoothed by convolution with the same Gaussian kernel with standard deviation 
$\sigma_{\rm sm}^\fkz$. 
For $\fkz(z)$, the flow speed must be slower than for $\visc(z)$ to remain in the linear regime, especially for the more hydrophobic surfaces (see Section \ref{sec:VeloFitsLinResp}).
For each system, the driving stress $F_a^\visc$ applied for extracting $\visc(z)$ is chosen to be four times greater than $F_a^\fkz$, the applied driving stress for extracting the friction coefficient for that system.
Therefore, more data is required to obtain comparable $f_f(z)$ profiles, and $f_f(z)$ benefits more from noise reduction here than in the case of the $\visc(z)$ calculation. 
Profiles $\fkz(z)$ are shown for various degrees of smoothing, as parametrized by $\sigma_{\rm sm}^\fkz$. 
The $\fkz(z)$ data for $\sigma_{\rm sm}^\fkz = 0.1$ {\AA}, shown here as a red curve, are those reproduced in the main text and used in subsequent modeling and analysis.

Similarly, \fig~\ref{fig:9profiles_gen_visc} shows, for all systems, $u(z)$ and $f_f(z)$ data from similar, but faster, driven-flow simulations, used to extract the viscosity profiles $\visc(z)$, which are also shown in the lower panels of the subfigures, directly below the corresponding raw data. 
Here, after the fitting procedure illustrated in \fig~\ref{fig:9_velo_fits}, just $u(z)$ is smoothed by convolution with a Gaussian kernel with standard deviation $\sigma_{\rm sm}^\visc$.
No data processing is applied to the input friction force profile $f_f(z)$ for calculating $\visc(z)$.

Smoothing of $u(z)$ is more important for calculation of $\visc(z)$, because $\visc(z)$ depends on the shear $\partial_z u(z)$, which is sensitively dependent on noise in the $u(z)$ data. 
Profiles $\visc(z)$ are shown for various degrees of smoothing, as parametrized by $\sigma_{\rm sm}^\visc$. 
The data for $\sigma_{\rm sm}^\visc = 0.3$ {\AA},  shown here as a blue curve, are those reproduced in the main text and used in subsequent modeling and analysis.
Here, it is apparent how the sharpness and height of the peaks in $\visc(z)$, especially in the first hydration layer, is attenuated by the smoothing.

In each panel of \figs~\ref{fig:9profiles_gen_fric} and \ref{fig:9profiles_gen_visc}, the corresponding mass density profiles for the liquid and surface, scaled by an arbitrary factor, are shown for use as a positional reference.
Also, the exponential fitting functions from \fig~\ref{fig:9_velo_fits} are shown over  just the domain where the they replace the $u(z)$ data before smoothing, for calculation of $\fkz(z)$ or $\visc(z)$.

\section{Numerically Solving the Stokes Equation}
\label{sec:NumStokes}
Consider a liquid layer bounded below by a flat solid surface and exposed above to a vapor phase, with both the solid--liquid and liquid--vapor interfaces having surface normal $\hat{z}$.
Let a volume element of the liquid be acted upon by two force densities: a friction force density $f_f(z) = -\fkz(z) u(z)$ and an applied driving force density $f_a(z)$.
In a steady state, the flow is governed by \eq~{\mainEqVeloReconst} from the main text,
\begin{equation*}
	\fkz(z) u(z) = f_a(z) + \partial_z \visc(z) u^\prime(z) 
	\,,
\end{equation*}
where $u(z)$ is the velocity, $\fkz(z)$ is the solid--liquid friction coefficient, $\visc(z)$ is the liquid shear viscosity, and $f^\prime(z)$ indicates the derivative of $f(z)$ with respect to $z$.
The product rule gives,
\begin{equation}
	f_a(z) = \fkz(z) u(z) - \visc^\prime(z) u^\prime(z) -  \visc(z) u^{\prime\prime}(z)
	\,.
\end{equation}
Let the space along $z$ be discretized into $N$ points with interval $h$ between them. 
We wish to solve for $u(z)$ in the discretized space, so we approximate the derivatives of $u(z)$ using central difference quotients,
\begin{align}
	f_a(z) \approx\; &\fkz(z) u(z) -  \visc^\prime(z) \frac{u(z+h) - u(z-h)}{2 h} 
\notag \\	
	&- \visc(z) \frac{u(z+h) -2u(z) + u(z-h)}{h^2}
	\,.
\end{align}
Indexing the discretization points by $i$, and letting $f_i \equiv f(z_i)$ allows this to be written more compactly as
\begin{equation}
	f_i^a \approx \fkz_i u_i -  \visc_i^\prime \frac{u_{i+1} - u_{i-1}}{2 h} 
	- \visc_i \frac{u_{i+1} -2u_i + u_{i-1}}{h^2}
	\,,
\end{equation}
where $f_i^a \equiv f_a(z_i)$. Grouping terms of $u_i$, $u_{i+1}$ and $u_{i-1}$ gives
\begin{align}
	f_i^a 
	\approx 
	&\left( \fkz_i + \frac{2\eta_i}{h^2} \right) u_i 
	\notag\\	
	&+ 
	\left( \frac{\eta_i^\prime}{2h} - \frac{\eta_i}{h^2} \right) u_{i+1}
	\notag\\		
	&+ 
	\left(-\frac{\eta_i^\prime}{2h} - \frac{\eta_i}{h^2} \right) u_{i-1} \,,
	\label{eq:Stokes_pre_matrix}
\end{align}
which is a set of $N$ coupled linear equations. 
Here, $\eta^\prime$ is calculated numerically from $\eta$ using a central difference scheme.
Thus, only $u$ is unknown.
This can be written in terms of the matrix $G$ as
\begin{equation}
	\boldsymbol{f}^a \approx G \boldsymbol{u} \,.
	\label{eq:Stokes_matrix_eq}
\end{equation}
See that \eq~\eqr{eq:Stokes_matrix_eq} can in principle be solved numerically for $\boldsymbol{u}$ in one step using a linear solver.
Taking $i\in\lbrace 1,\ldots, N \rbrace$, $u_1$ and $u_N$ are also functions of $u_0$ and $u_{N+1}$, i.e., two boundary conditions (BCs) are also needed.
Let the left (here, solid--liquid) and right (here, liquid--vapor) BCs be given by matrices $G^{ L}$ and $G^{ R}$ such that $G=G^0+G^{ L}+G^{ R}$, with
\begin{align}
	&G^0_{ii} = \fkz_i + \frac{2\eta_i}{h^2} \,, \notag\\	
	&G^0_{i(i+1)} = \frac{\eta_i^\prime}{2h} - \frac{\eta_i}{h^2} \,, \notag\\		
	&G^0_{i(i-1)} = -\frac{\eta_i^\prime}{2h} - \frac{\eta_i}{h^2} \,, \notag\\		
	&G^0_{ij} = 0 \quad \text{for} \quad j \notin [i, i+1, i-1]	\,.
	\label{eq:Stokes_matrix_no_BC}
\end{align}
As defined, $G^0\boldsymbol{u} $ contains all terms of the R.H.S.\ of \eq~\eqr{eq:Stokes_pre_matrix} except the boundary terms proportional to $u_0$ and $u_{N+1}$.
Thus, taking $G^{ L}={\bf 0}$ is equivalent to assuming that $u_0=0$, i.e., it amounts to a zero-velocity BC.
In addition to the zero-velocity BC, another obvious choice is a no-shear BC, where the derivative of the velocity vanishes.
For the (left) solid--liquid boundary at $u_1$, letting $u_1^\prime = 0$ and approximating $u_1^\prime$ with the backward-difference formula gives the condition $u_0=u_1$, and 
\begin{equation}
	G_{ij}^{L} = \delta_{i1}\delta_{1j} 
	\left(
	-\frac{\eta_1^\prime}{2h} - \frac{\eta_1}{h^2}
	\right)\,,
	\label{eq:Stokes_matrix_L_BC}
\end{equation}
where $\delta_{ij}$ is the Kronecker delta.
For the (right) liquid--vapor boundary, the no-shear BC can be obtained by a similar procedure, except that the forward-difference formula is used, and the added term is that for $u_{N+1}$, i.e.,
\begin{equation}
	G_{ij}^{R} = \delta_{iN}\delta_{Nj}
	\left( 
	\frac{\eta_N^\prime}{2h} - \frac{\eta_N}{h^2} 
	\right) \,.
	\label{eq:Stokes_matrix_R_BC}
\end{equation}
In the case of the solid--liquid--vapor systems studied in this work, it is reasonable to take a zero-velocity BC on the left, near the solid--liquid interface ($G^{ L}={\bf 0}$), and a no-shear BC on the right, near the liquid--vapor interface ($G^{ R}$ given by \eq~\eqr{eq:Stokes_matrix_R_BC}). 
These are the boundary conditions assumed wherever \eq~{\mainEqVeloReconst} is used to model velocity profiles, except in Section \ref{sec:NaiveVeloModels}, where both zero-velocity \textit{and} no-shear boundary conditions are explored at the solid--liquid interface.
 
\section{Verifying Linear-Response Regime by Velocity Profile Fits}
\label{sec:VeloFitsLinResp}

\begin{figure*}[!t]
	\centering
	\includegraphics{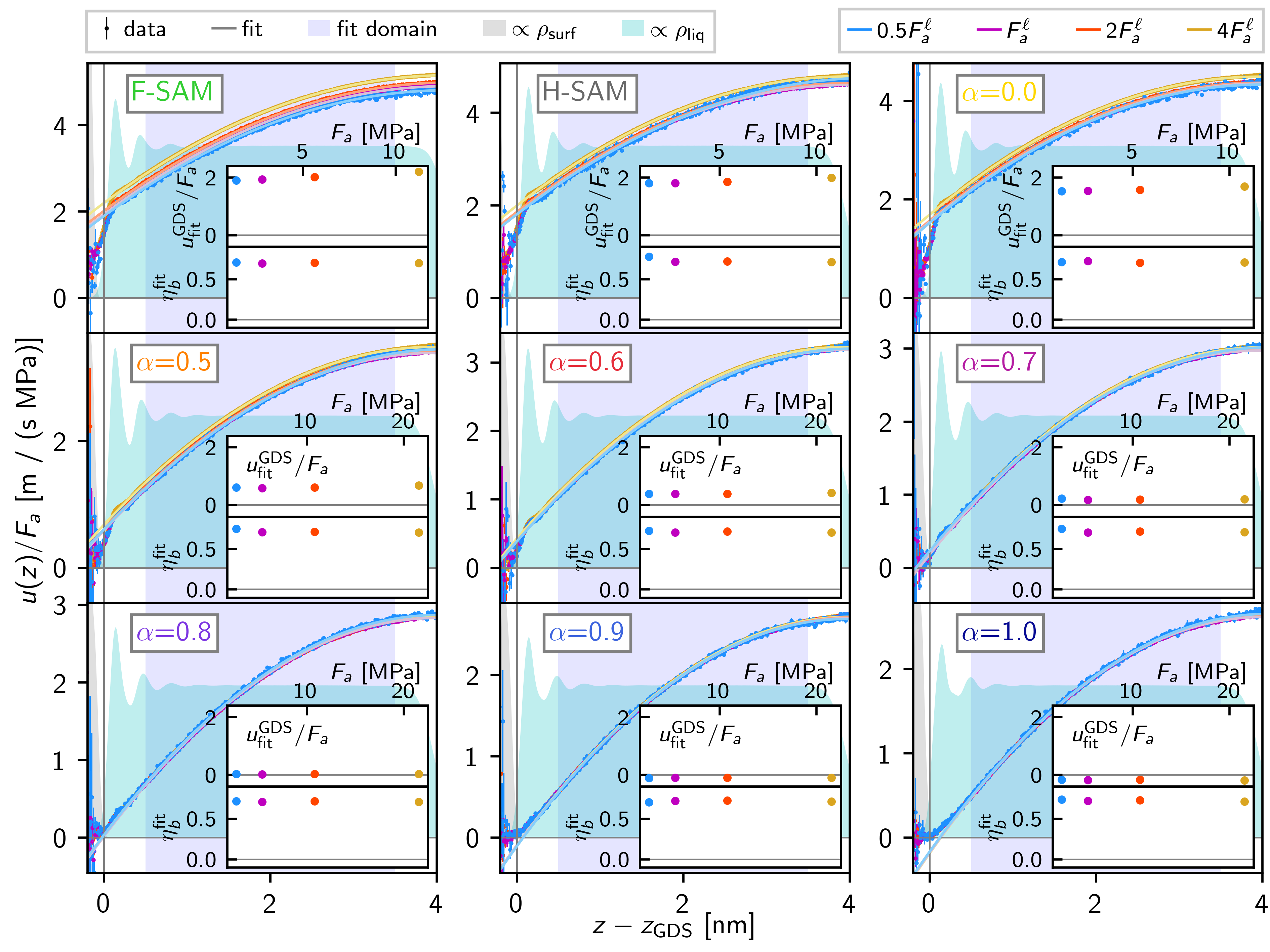}
	\caption{
	Velocity divided by applied driving stress $u(z) / F_a$ for different driving stresses shown alongside quadratic fits $u_{\rm fit}(z) / F_a$, where $u_{\rm fit}(z) $ is obtained by fitting \eq~\eqr{eq:quadratic_fit_fn} to  $u(z)$ in the bulk domain, which is indicated by the blue shaded area.
	The water and SAM densities are also shown on an arbitrary scale as shaded cyan and gray areas, respectively.
	The upper insets plot $u_{\rm fit}(z_{\rm GDS}) / F_a$, in units of \mbox{m$\,$/$\,$(s$\,$MPa)}, over $F_a$.
	The lower insets show the bulk viscosity, in units of \mbox{mPa$\,$s}, calculated from the curvature of the fits via \eq~\eqr{eq:visc_calc_from_fit}, over $F_a$.
	\label{fig:9_normvelos}}
\end{figure*}

To check if a system is in the linear-response regime for both the surface friction and viscosity, the quadratic function 
\begin{equation}
	u_{\rm fit}(z) = A \left(z-z_{\rm GDS}\right)^2 + B\left(z-z_{\rm GDS}\right) + C \,,
	\label{eq:quadratic_fit_fn}
\end{equation}
is fitted to the bulk-domain velocity profiles $u(z)$ of driven-flow NEMD simulations at four different driving stresses: $0.5 F_a^\fkz$, $F_a^\fkz$, $2 F_a^\fkz$, and $4 F_a^\fkz = F_a^\visc$, where $F_a^\fkz$ is the driving stress applied to extract $\fkz(z)$, and $F_a^\visc$  the driving stress applied to extract $\visc(z)$, for that respective system.
From the fit of \eq~\eqr{eq:quadratic_fit_fn}, two quantities can be immediately calculated:
the first is the viscosity in the bulk, given by
\begin{equation}
	\visc_b^{\rm fit} = -\frac{ f_a(z_b)}{2 A} \,,
	\label{eq:visc_calc_from_fit}
\end{equation}
where $f_a(z_b)$ is the driving force density in the fitted domain, and the second is the effective slip velocity at $z_{\rm GDS}$, given by $C$.

\fig~\ref{fig:9_normvelos} plots the velocity profile $u(z)$ divided by the total applied driving stress $F_a$, for each of the four driving stresses and for each system studied in this work.
In the linear-response regime of both viscosity and friction, $u(z)/F_a$ should collapse onto a master curve for all driving stresses $F_a$.
Outside the linear-response regime of the viscosity, the curvature of $u(z)/F_a$ should disagree  for different values of $F_a$, while outside the linear-response  regime of the surface--liquid friction, $u(z)/F_a$ should disagree by a constant $y$-offset for different values of $F_a$.
Also plotted in \fig~\ref{fig:9_normvelos} are  $u_{\rm fit}(z) / F_a$, where  $u_{\rm fit}(z)$ are the corresponding fits of \eq~\eqr{eq:quadratic_fit_fn}.
In each panel of \fig~\ref{fig:9_normvelos}, the upper inset shows $u_{\rm fit}(z_{\rm GDS})/F_a$ plotted over $F_a$, the lower inset, $\visc_b^{\rm fit}$ (calculated via \eq~\eqr{eq:visc_calc_from_fit}) over $F_a$.

The data for the most hydrophilic systems agree very well across different values of $F_a$ indicating that the system is in the linear regime for both friction and viscosity there.
Moving to the more hydrophobic systems, $u(z)/F_a$ for the largest driving stress,  \mbox{$F_a = 4F_a^\fkz$}, is increasingly shifted along the $y$-axis, above the other lines.
For the F-SAM/water system, the data for both  the $F_a = 2F_a^\fkz$ and $4F_a^\fkz$ are shifted, indicating that the slip is in the linear regime only when $F_a \le F_a^\fkz$. 
Thus, we conclude that the friction-coeffient profiles $\fkz(z)$ are all extracted in the linear regime of surface--liquid friction.

The viscosity data on the other hand, agree well across $F_a$, with the exception, for some systems, of the data from smallest driving stresses, $F_a = 0.5F_a^\fkz$, which appear to disagree mostly due to noise in the velocity data.
Thus, we conclude that the viscosity profiles $\visc(z)$ are also all extracted in the linear regime of the bulk viscosity.
\begin{figure*}[!ht]
	\centering
	\includegraphics{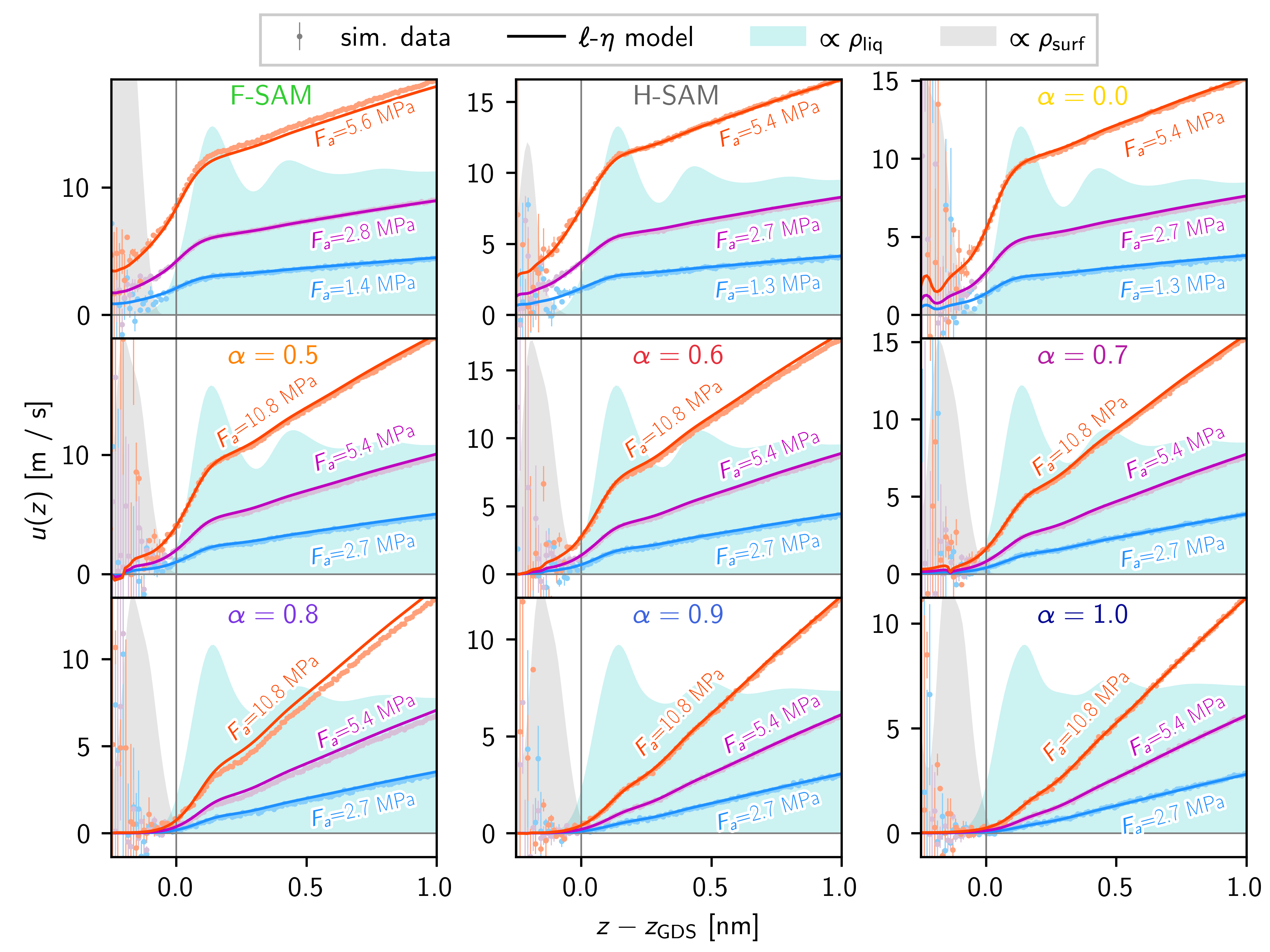}
	\caption{
	Velocity profiles $u(z)$ extracted from driven-flow simulations with different applied driving stresses $F_a$, compared with those calculated using the $\fkz$-$\visc$ model, i.e., by solving \eq~{\mainEqVeloReconst} from the main text numerically.
	\label{fig:9velo_reconst}}
\end{figure*}

\section{Modeling Flow for All \\ Systems}
\label{sec:VeloReconst}

For each system, driven-flow NEMD simulations are carried out at three applied driving stresses $F_a$.
These are the driving stress $F_a^\fkz$, used for the extraction of $\fkz(z)$ for the respective system (as shown in \fig~\ref{fig:9profiles_gen_fric}), as well as $0.5 \times F_a^\fkz$ and $2 \times F_a^\fkz$.
In \fig~\ref{fig:9velo_reconst}, velocity profiles $u(z)$ from each of these simulations are compared with profiles calculated using the $\fkz$--$\visc$ model, i.e., by numerically solving \eq~{\mainEqVeloReconst} from the main text, using the $\fkz(z)$ and $\visc(z)$ profiles plotted in \fig~{\mainFigAllProfs} in the main text, and taking the applied stress to be
\begin{equation}
	f_a(z) = F_a \frac{\rho_{\rm liq}(z)}{\int \mathrm{d}z^\prime \, \rho_{\rm liq}(z^\prime)}\,.
\end{equation}
The modeled velocity profiles agree remarkably well with the simulation data, which validates our approach, notably the locality of $\fkz$ and $\visc$, as mentioned in Sections \ref{sec:PosDepFric} and \ref{sec:NavierStokes} and the main text, and indicates that $\fkz$ and $\visc$ were extracted in the linear regime.


\begin{figure*}[!ht]
	\centering
	\includegraphics{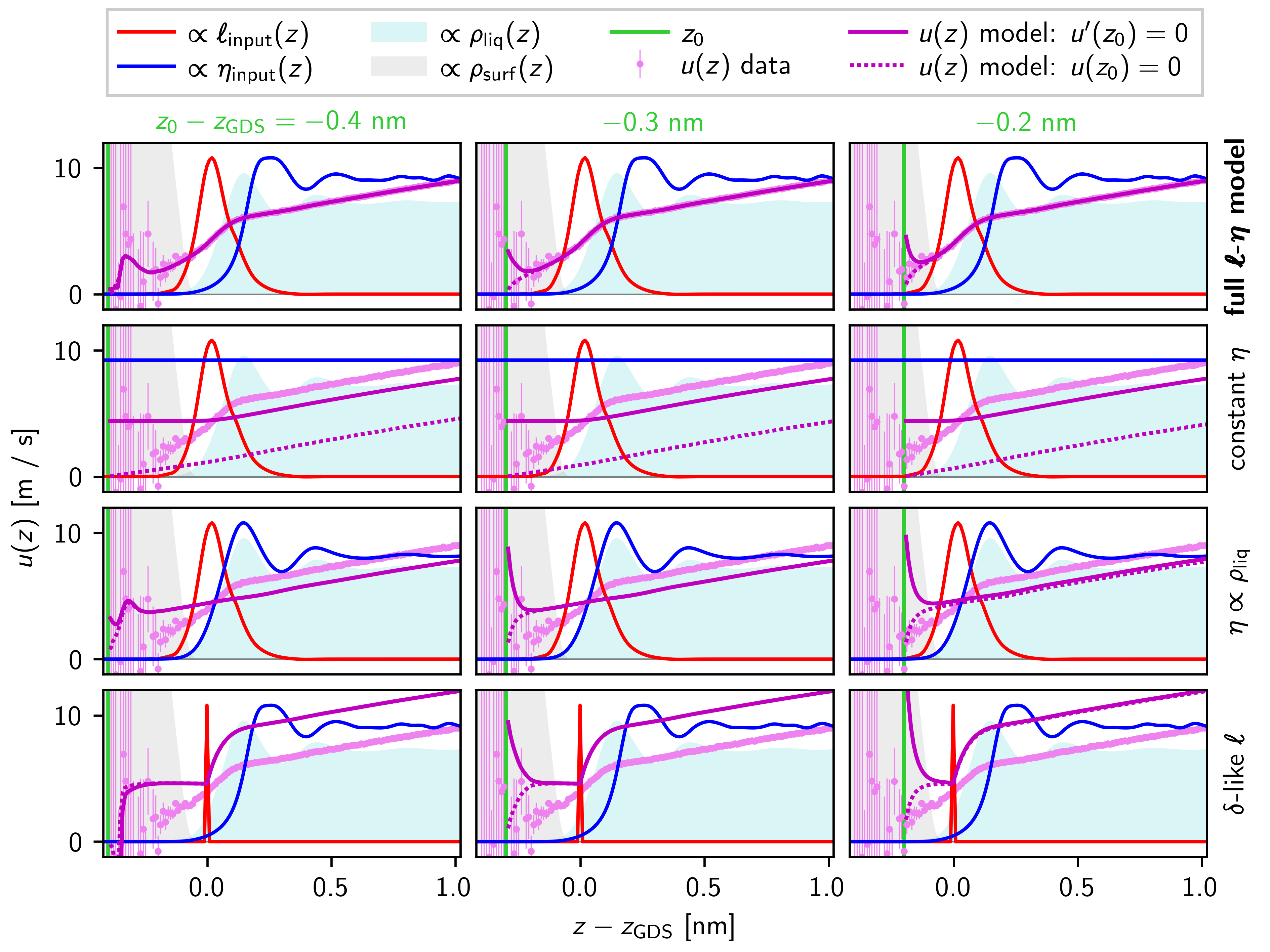}
	\caption{
	Comparison of different approaches to modeling the velocity profile of water on an F-SAM to a velocity profile extracted directly from a driven-flow NEMD simulation with a driving stress of 2.8~MPa (plotted as dots with error bars).
	Each row shows data for a different model (see the labels to the right of each row);
	the models are parametrized by the input surface--liquid friction coefficient profiles $\fkz_{\rm input}(z)$ and viscosity profiles $\visc_{\rm input}(z)$, which are also plotted with an arbitrary scaling factor.
	Each column shows data for a different lower domain boundary $z_0$ (see the labels at the top of each column).
	In each panel, the velocity profile is modeled with two different boundary conditions, $u^\prime(z_0) = 0$ (solid line) and $u(z_0) = 0$ (dotted line).
	The mass-density profiles of the F-SAM and water are also shown as shaded areas as a positional reference with an arbitrary  scaling factor.
	\label{fig:12velo_reconst_cf}}
\end{figure*}

\section{Other Approaches to \\ Modeling Flow}
\label{sec:NaiveVeloModels}
We show in \fig~\ref{fig:9velo_reconst} that the $\fkz$--$\visc$ model, i.e., solving \eq~{\mainEqVeloReconst} from the main text numerically using the extracted $\fkz(z)$ and $\visc(z)$ profiles, is accurate for modeling nanoscopic flows near surfaces.
This leaves open the question as to whether simpler approaches would predict the behavior equally as well.
This is explored in \fig~\ref{fig:12velo_reconst_cf}, where each panel plots the velocity profile $u(z)$ extracted from an F-SAM/water driven-flow simulation with driving stress $F_a = 2.8$ MPa, compared to a flow generated by a numerical model.
Each row shows results from a different model, i.e, a solution of \eq~{\mainEqVeloReconst} from the main text with different simplifying assumptions about the input profiles $\fkz_{\rm input}(z)$ and $\visc_{\rm input}(z)$.
A finite domain along $z$ must be chosen in which to solve the equation. 
We call the lower bound of this domain $z_0$, and the columns of \fig~\ref{fig:12velo_reconst_cf} correspond to three choices of  $z_0$, all of which are below where $\fkz(z)$ and $\visc(z)$ have decayed to near zero.
We plot the data for different values of $z_0$ to check whether the modeled flow is dependent on $z_0$ or not.
In each panel, there are two modeled velocity profiles plotted, a solid and  a dotted line, which correspond to two boundary conditions: $u^\prime(z_0) = 0$ (no shear) and  $u(z_0) = 0$ (zero velocity), respectively.
At the upper boundary at $z-z_{\rm GDS} = z_1 \approx 4$ nm, the no-shear boundary condition, $u^\prime(z_1) = 0$, is always used.

The top row shows our complete $\fkz$--$\visc$ model, which captures the velocity profile very accurately, for all values of $z_0$ and for both boundary conditions. There is slight disagreement near $z_0$, but this is in a zone with effectively zero liquid density and poor statistics.

The second row shows a simplified model where the shear viscosity is assumed to be constant everywhere.
The value taken is the shear viscosity from the bulk, $\visc_b$.
This model fails to capture both the detailed interfacial behavior and the net slip of the liquid, reflected as a constant shift in velocity along the $y$-direction.
For the $u(z_0)=0$ boundary condition, the slip is also dependent on the choice of $z_0$.

The third row shows a simplified model where the viscosity is assumed to be proportional to the density of water molecules, scaled such that it approaches the correct viscosity in the bulk, $\visc_b$.
This model fails very similarly to the constant viscosity model in the row above, though it seems to be independent of $z_0$.

The bottom row shows a model where only the Navier friction coefficient $\fks$, rather than the entire $\fkz(z)$ profile, is used. 
Here, the arbitrary assumption that the friction acts at the Gibbs dividing surface of the liquid is made. 
Like the complete $\fkz$--$\visc$ model, this model seems to be independent of $z_0$, but it again fails to capture both the detailed behavior near the interface and more importantly, the slip behavior. 

Thus, when modeling flows with the Stokes equation, accounting for the position-dependence of both the surface-friction and the viscosity, as the complete $\fkz$--$\visc$ model does, gives far more accurate results on sub-nanometer scales.

\section{Contact Angles}

\begin{figure}[!h]
	\centering
	\includegraphics[width=5cm]{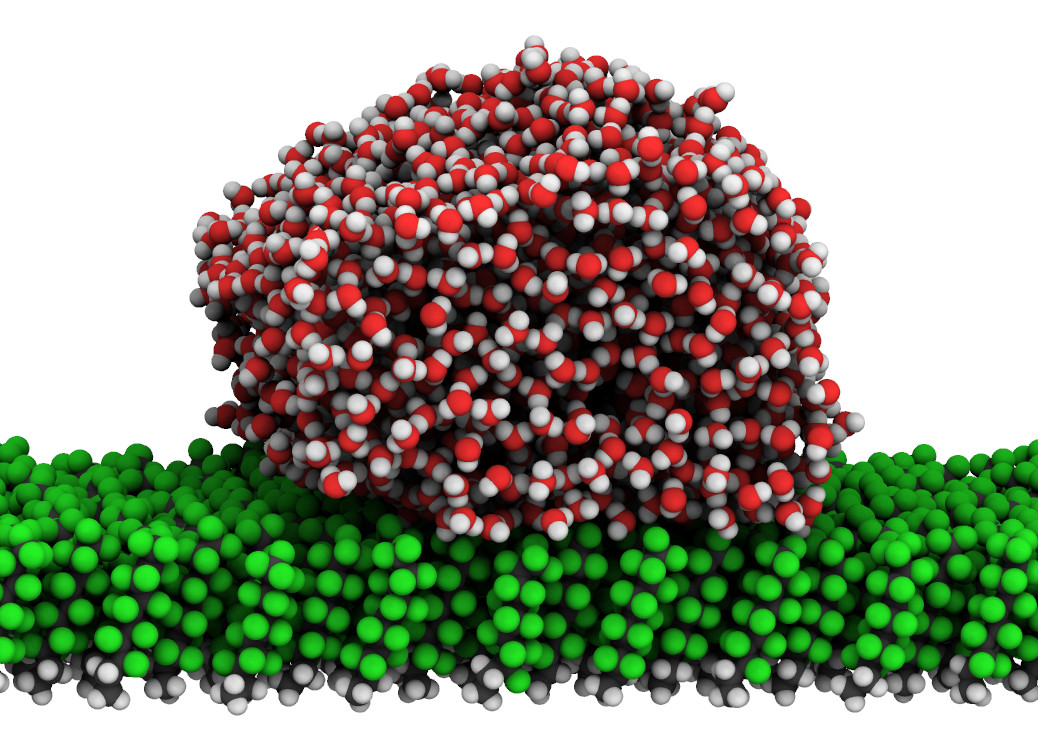}
	\vspace*{2mm}
	\includegraphics{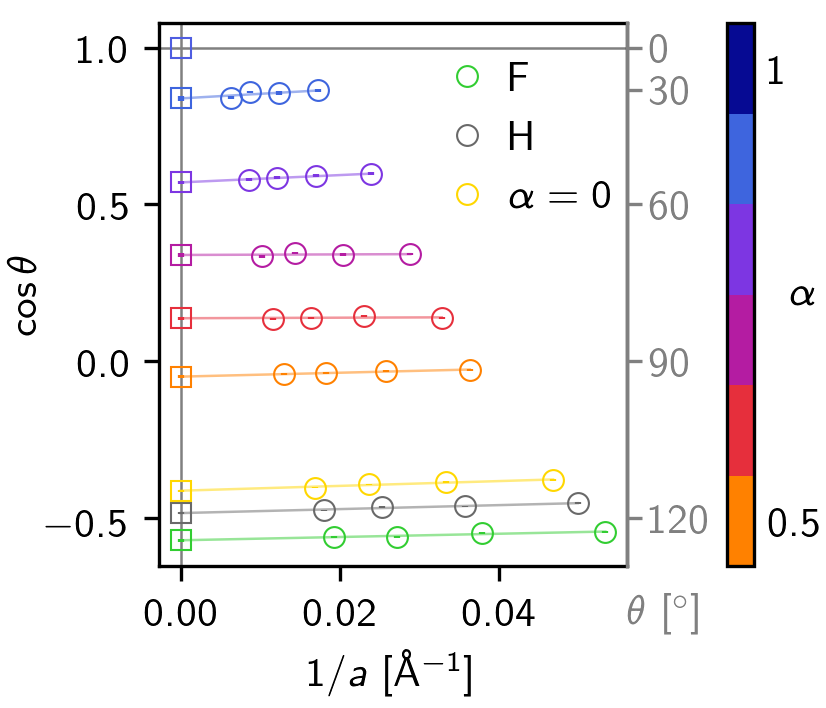}
	\caption{Extrapolation of microscopic contact angles to the macroscopic limit. 
	Here, $a$ is the droplet footprint radius.
	The circles indicate microscopic contact angles extracted from cylindrical droplet simulations, and the squares indicate the macroscopic value obtained by extrapolation to $1/a=0$ of the linear fits, which are shown as solid lines.
	Shown above the plot is a snapshot of a cylindrical droplet of 2048 water molecules on an F-SAM (the smallest simulated droplet).
	\label{fig:CAextrap}}
\end{figure}

We measure the contact angles of all of our systems by simulating cylindrical droplets, which are continuous over the periodic boundary along the $y$-direction.
Cylindrical droplets have been shown to be less susceptible to finite size effects than spherical ones, and have the same contact angle in the macroscopic limit \cite{2018_Kanduc_PRE}.
\fig~\ref{fig:CAextrap} shows a snaphot of a 2048-molecule cylindrical droplet on an F-SAM.
For each surface type, we simulate droplets of 2048, 4096, 8192, and 16384 water molecules and extract contact angles $\theta_\mu$ from these.
These ``microscopic'' contact angles  $\theta_\mu$ can deviate from the macroscopic contact angle $\theta$ due to finite-size effects. 
Thus, assuming a first-order correction to the work of adhesion,  
\begin{equation}
	\cos \theta_\mu = \cos \theta - \frac{C_0}{a} \,,
\end{equation} 
where $a$ is the footprint radius of the droplet, we extrapolate to the $1/a\to 0$ limit.
In the case of the cylindrical droplets, $a$ is half the footprint length along the $x$-direction. 
Extrapolations for all systems studied are shown in \fig~\ref{fig:CAextrap}.
For the ($\alpha$=1)-SAM, the water fully wetted the SAM, so no contact angle could be extracted. 

The method we used for extracting contact angles was first published in Ref.~\refcite{2024_Carlson_JPCL} and we outline it briefly here.
From each simulation, the mass-density profile $\rho_{\rm drop}(z)$ along the surface normal $\hat{z}$ of the liquid is extracted. 
The mass density of a planar liquid slab adsorbed on the same surface, $\rho_{\rm slab}(z)$, is also extracted from an equilibrium SAM/water-slab simulation.
The droplet geometry can then be obtained by fitting the function 
\begin{align}
	&\frac{\rho_{\rm drop}(z)}{\rho_{\rm slab}(z)}
	= 
	\frac{1}{2 L_x} \times 
	\notag \\
	&\int_{0}^{L_x} \mathrm{d}x \, 
	\left( 
		1 - \tanh	
		\left( 
			\frac{\sqrt{x^2 + (z-z_0)^2} - R_0}{d} 
		\right)
	\right)\,
	\label{eq:1d_droplet_mtd_fit}
\end{align}
over the fitting parameters $z_0$, $R_0$, and $d$, which are the center-position of the droplet, droplet radius, and the sigmoid width $d$.
Here, $L_x$ is the system box length along the $x$-direction. 
The integral in \eq~\eqr{eq:1d_droplet_mtd_fit} is evaluated numerically.
The droplet radius $R_0$ is in fact the radius to the half-maximum position of the sigmoid, whereas the desired radius is the radius to the Gibbs dividing surface.
These are the same for a flat interface, but not for a curved one.
The Gibbs radius can be calculated for a cylindrical droplet via 
\begin{equation}
	R_{\rm Gibbs}
	= 
	d
	\sqrt{\frac{-1}{2} \mathrm{Li}_2\left( e^{-2R_0/d}\right)} \,,
\end{equation}
where $\mathrm{Li}_2$ is the polylogarithm function.
We approximate this by the Taylor expansion up to second order,
\begin{equation}
	R_{\rm Gibbs}
	\approx
	R_0 \left(
	1 + \frac{\pi^2}{24} 
	\left(
		\frac{d}{R_0}
	\right)^2
	\right) \,.
\end{equation}
From the droplet center position $z_0$, the Gibbs radius $R_{\rm Gibbs}$, and the position of the Gibbs dividing surface at the bottom of the planar liquid $z_{\rm Gibbs}$, the contact angle can be calculated from
\begin{equation}
	\cos \theta =\frac{z_{\rm Gibbs} - z_0}{R_{\rm Gibbs}} \,.
\end{equation}
A more detailed derivation and discussion can be found in Ref.~\refcite{2024_Carlson_JPCL} and its supplement.


\section{Effective Viscosity Profiles}
\label{sec:EffVisc}
Consider a system with a surface--liquid interface parallel to the $xy$-plane, where the liquid is driven along $\hat{x}$, as described in Sections \ref{sec:PosDepFric} and \ref{sec:PosDepVisc}. For a steady state flow, the liquid is governed by \eq~\eqr{eq:stokes_ss} from Section \ref{sec:PosDepVisc}, which reads
\begin{equation*}
	f(z)
	=
	- \partial_z \visc(z) \partial_z u(z) \,,
\end{equation*}
where  $u(z)$ is the velocity profile of the liquid and $f(z)$ is the external force density on the liquid, which may be written $f(z) = f_f(z) + f_a(z)$, where $f_f(z)$ is the surface--liquid friction force density and $f_a(z)$ is the applied driving force density.
If $f(z)$ and $u(z)$ are obtained to a sufficiently high degree of accuracy, the shear viscosity profile $\eta(z)$ calculated via \eq~\eqr{eq:visc_prof_calc} should be found to plateau at a constant value $\eta_b$ in the bulk of the liquid far from the interface. 
Once $\eta_b$ is known, \eq~\eqr{eq:stokes_ss} may be integrated over $z$ to give
\begin{figure}[!t]
	\centering
	\includegraphics{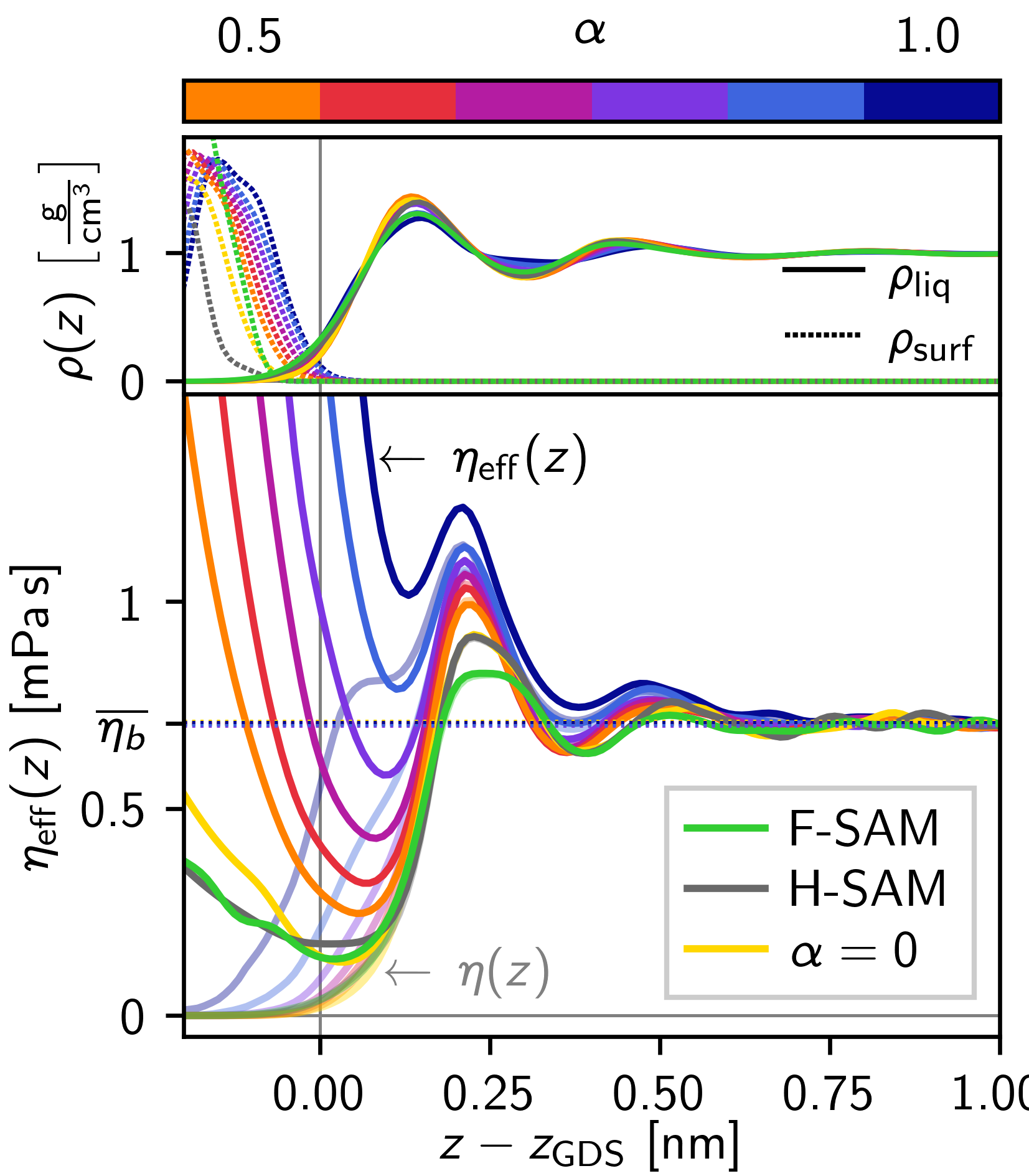}
	\caption{
	Effective viscosity profiles calculated via \eq~\eqr{eq:visc_prof_calc_eff} for all systems studied in this work. 
	Also plotted as translucent lines for comparison are the viscosity profiles calculated via \eq~\eqr{eq:visc_prof_calc}.
	\label{fig:all_visc_profiles_cf_2panel_effective}}
\end{figure}
\begin{align}
	\int_{z}^{z_b} \mathrm{d}z^\prime \, f(z^\prime)
	&=
	- \int_{z}^{z_b} \mathrm{d}z^\prime \,
	\partial_{z^\prime} \eta(z^\prime) \partial_{z^\prime} u(z^\prime) \notag \\
	&=
	\eta(z) \partial_{z} u(z) - \eta_b \partial_{z_b} u(z_b) \notag \\
	\Rightarrow \:\: 
	\eta(z) = \frac{1}{\partial_{z} u(z)} 
	&\left(
		\eta_b \partial_{z_b} u(z_b) +
		\int_{z}^{z_b} \mathrm{d}z^\prime \, f(z^\prime)
	\right)
	 \,. \label{eq:int_stokes_ss_reverse}
\end{align}
Equation~\eqr{eq:int_stokes_ss_reverse} is also useful for calculating the effective viscosity profile $\eta_{\rm eff}(z)$, where position-dependent surface--liquid friction forces $f_f(z)$ are ignored, with their effects instead being attributed to local variations in the liquid viscosity;
\begin{align}
	\eta_{\rm eff}(z) = \frac{1}{\partial_{z} u(z)} 
	&\left(
		\eta_b \partial_{z_b} u(z_b) +
		\int_{z}^{z_b} \mathrm{d}z^\prime \, f_a(z^\prime)
	\right)
	 \,.
	\label{eq:visc_prof_calc_eff}
\end{align}
For a steady-state Couette flow between oppositely moving parallel plates, where $f_a(z) = 0$ in the liquid, \eq~\eqr{eq:visc_prof_calc_eff} reduces to
\begin{equation}
	\eta_{\rm eff}(z) = \frac{\eta_b \partial_{z_b} u(z_b)}{\partial_{z} u(z)} =  \frac{F_a}{\partial_{z} u(z)} 
	 \,,
	\label{eq:visc_prof_calc_eff_no_local_fa}
\end{equation}
where $F_a$ is the total stress from the driving force. 
This formulation has been used in previous works \cite{2013_Bonthuis_JPCB, 2017_Schlaich_NL, 2021_WoldeKidan_Langmuir}.

\fig~\ref{fig:all_visc_profiles_cf_2panel_effective} shows effective viscosity profiles $\visc_{\rm eff}$,  calculated via \eq~\eqr{eq:visc_prof_calc_eff}, for all systems studied in this work.
The profiles diverge near the interface to account for the rapid decrease in flow there, which is in fact due to surface--liquid friction.
This divergence in the effective viscosity profile has also be seen in previous works \cite{2013_Bonthuis_JPCB, 2017_Schlaich_NL, 2021_WoldeKidan_Langmuir}. 

\section{Bulk Shear Viscosity from the Green--Kubo Relation}
\label{sec:GKVisc}

In the linear-response regime, the Green--Kubo relation,
\begin{equation}
	\widetilde{\eta}(\nu) = \beta V \int_0^{\infty} \mathrm{d}t \,  
	e^{-i 2 \pi \nu t}
	\left\langle
		P_{ij}(t) P_{ij}(0)
	\right\rangle\,,
	\label{eq:GreenKubo_single}
\end{equation}
gives the frequency-dependent viscosity $\widetilde{\eta}(\nu)$ from a simulation of bulk liquid with periodic boundary conditions, where  $P_{ij}(t)$ is any off-diagonal element of the stress tensor ($i\neq j$), $\beta= 1/(k_B T)$ is the inverse thermal energy, and $V$ is the system volume \cite{2009_Gonzalez_JCP, 2020_Schulz_PRF}.
Here, $\left\langle \ldots \right\rangle$ indicates a thermal average, and $\left\langle P_{ij}(t) P_{ij}(0) \right\rangle$ is the time autocorrelation function (ACF) of $P_{ij}(t)$.
However, \textit{all} elements of $P_{ij}$ contain information about the shear viscosity, so \eq~\eqr{eq:GreenKubo_single} makes suboptimal use of the existing data for obtaining $\widetilde{\eta}(\nu)$. 
Let $\Pi_{ij}(t)$ be the deviatoric stress tensor,
\begin{equation}
	\Pi_{ij}(t) = P_{ij}(t) - p(t) \,,
\end{equation}
where $p$ is the time-dependent pressure,
\begin{equation}
	p(t) = \frac{1}{3} \sum_{i=1}^3 P_{ii}(t) \,.
\end{equation}
See that $\Pi_{ij}(t)$ is trace free by construction, i.e.,
\begin{equation}
	\Pi_{xx}(t) + \Pi_{yy}(t) +  \Pi_{zz}(t) = 0\,.
	\label{eq:tracelessness}
\end{equation}
The volume viscosity, which parameterizes dissipation under compression, can be found via fluctuations of the pressure $p(t)$, while the shear viscosity, which parameterizes dissipation under shear, can be found via fluctuations of the nine elements of $\Pi_{ij}(t)$. 
Because $\Pi_{ij}(t)$ is symmetric, the number of independent elements is reduced by three.
Equation~\eqr{eq:tracelessness} imposes a further constraint, reducing the total number of independent elements to five. These are conventionally the off-diagonal elements $\Pi_{xy}$, $\Pi_{yz}$ and $\Pi_{xz}$, and two additional stresses constructed  from the diagonal elements, \cite{1998_Alfe_PRL}
\begin{align}
	&Q_{xy}=\frac{\Pi_{xx} - \Pi_{yy}}{2} \quad \text{and}  \notag \\
	&Q_{yz} =\frac{\Pi_{yy} - \Pi_{zz}}{2} \,.
\end{align}
Using the tracelessness given by \eq~\eqr{eq:tracelessness}, it is easy to show that the diagonal elements can all be expressed in terms of $Q_{xy}$ and $Q_{yz}$,
\begin{align*}
	&\Pi_{xx} = \frac{4 Q_{xy} + 2 Q_{yz}}{3} \,,  \\
	&\Pi_{yy} = \frac{-2 Q_{xy} + 2 Q_{yz}}{3} \,,  \\
	&\Pi_{zz} = \frac{-2 Q_{xy} -4 Q_{yz}}{3} \,,
\end{align*}
i.e., $Q_{xy}$ and $Q_{yz}$ span the diagonal elements.
The elements $Q_{xy}$ and $Q_{yz}$ are themselves off-diagonal elements of the stress tensor in frames of reference rotated by $45^\circ$. 
This can be expressed more concretely in terms of rotation matrices.
The matrices for rotation about $x$ and $z$ are given by  
\begin{align*}
	R_x(\theta) &= 
	\begin{pmatrix}
		1 & 0 & 0\\
		0 & \cos \theta  & \sin \theta\\ 
		0 & -\sin \theta  & \cos \theta\\ 
	\end{pmatrix}
	\,, \\
	R_z(\theta) &= 
	\begin{pmatrix}
		\cos \theta  & \sin \theta & 0\\ 
		-\sin \theta  & \cos \theta & 0\\ 
		0 & 0 & 1
	\end{pmatrix} \,.
\end{align*}
It is easy to show by matrix multiplication and using the symmetry of $\Pi$ that 
\begin{equation*}
	\left( R_z(-45^\circ) \cdot \Pi \cdot R_z^T(-45^\circ)\right)_{xy} = Q_{xy}  \,,
\end{equation*}
and 
\begin{equation*}
	\left( R_x(-45^\circ) \cdot \Pi \cdot R_x^T(-45^\circ)\right)_{yz} = Q_{yz}  \,.
\end{equation*}
Thus, for an isotropic system, $Q_{xy}$ and $Q_{yz}$ have identical ACFs to the off-diagonal elements of $\Pi_{ij}$.
Let  $Q_{ij}^t \equiv Q_{ij}(t)$ and $\Pi_{ij}^t \equiv \Pi_{ij}(t)$. The ACF of $Q_{ij}^t$ is given by
\begin{align}
	&\left\langle
		Q_{ij}^t Q_{ij}^0
	\right\rangle = 
	\frac{1}	{4}
	\left\langle
		(\Pi_{ii}^t - \Pi_{jj}^t)
		(\Pi_{ii}^0 - \Pi_{jj}^0))
	\right\rangle =
	\label{eq:Q_ACF_0}	
	\\
	&\frac{1}	{4}
	\left(
	\left\langle
		\Pi_{ii}^t \Pi_{ii}^0
	\right\rangle
	+
	\left\langle
		\Pi_{jj}^t \Pi_{jj}^0
	\right\rangle 
	-
	\left\langle
		\Pi_{ii}^t \Pi_{jj}^0
	\right\rangle
	-	
	\left\langle
		\Pi_{jj}^t \Pi_{ii}^0
	\right\rangle
	\right)
	\,.
	\notag
\end{align}
Because of isotropy,
\begin{align}
	&\left\langle 
		\Pi_{ii}^t \Pi_{ii}^0
	\right\rangle 
	= 
	\left\langle 
		\Pi_{jj}^t \Pi_{jj}^0
	\right\rangle  
	\quad
	\text{and}
	\\
	&\left\langle
		\Pi_{ii}^t \Pi_{jj}^0
	\right\rangle
	=
	\left\langle
		\Pi_{kk}^t \Pi_{ll}^0
	\right\rangle \,,
\end{align}
for any $i$, $j$, $k$, and $l$.
Thus, \eq~\eqr{eq:Q_ACF_0}	reduces to
\begin{equation}
	\left\langle
		Q_{ij}^t Q_{ij}^0
	\right\rangle = 
	\frac{1}{2}
	\left(
	\left\langle 
		\Pi_{ii}^t \Pi_{ii}^0
	\right\rangle 
	-
	\left\langle
		\Pi_{ii}^t \Pi_{jj}^0
	\right\rangle
	\right)\,.
	\label{eq:Q_ACF_1}	
\end{equation}
From the tracelessness of $\Pi_{ii}$, it follows that
\begin{align}
	\left\langle
		\Pi_{ii}^t \Pi_{jj}^0
	\right\rangle
	&= 
	\left\langle
		\Pi_{ii}^t
		\left( 
			-\Pi_{ii}^0 - \Pi_{kk}^t
		\right)
	\right\rangle \notag \\
	&= 
	-
	\left\langle
		\Pi_{ii}^t \Pi_{ii}^0
	\right\rangle
	-
	\left\langle
		\Pi_{ii}^t \Pi_{kk}^0
	\right\rangle \,,
\end{align}
where $i\neq j$, $j \neq k$, and $i\neq k$.
From isotropy,
\begin{align}
	\left\langle
		\Pi_{ii}^t \Pi_{jj}^0
	\right\rangle
	&= 
	-
	\left\langle
		\Pi_{ii}^t \Pi_{ii}^0
	\right\rangle
	-
	\left\langle
		\Pi_{ii}^t \Pi_{jj}^0
	\right\rangle  \notag \\
	\Rightarrow  \quad
	\left\langle
		\Pi_{ii}^t \Pi_{jj}^0
	\right\rangle
	&=
	-\frac{1}{2}
	\left\langle
		\Pi_{ii}^t \Pi_{ii}^0
	\right\rangle \,.
\end{align}
Thus, \eq~\eqr{eq:Q_ACF_1} gives
\begin{equation}
	4\left\langle
		Q_{ij}^t Q_{ij}^0
	\right\rangle = 
	3
	\left\langle 
		\Pi_{ii}^t \Pi_{ii}^0
	\right\rangle \,.
	\label{eq:Q_ACF_2}	
\end{equation}
We wish to apply \eq~\eqr{eq:GreenKubo_single} to each of $\Pi_{xy}$, $\Pi_{yz}$ and $\Pi_{xz}$, $Q_{xy}$ and $Q_{yz}$ and average the results.
To minimize the effect of numerical errors in   the pressure tensor elements from simulations, we average over ACFs of both off-diagonal elements, i.e., the average will include terms with ACFs of the elements $\Pi_{xy}$, $\Pi_{yx}$, $\Pi_{yz}$, $\Pi_{zy}$, $\Pi_{xz}$, and $\Pi_{zx}$.
Because we effectively double count ACFs of $\Pi_{ij}$ in the average, ACFs of $Q_{ij}$ should also be double counted so that they are equally weighted, i.e., there should be a total of four ACFs of $Q_{ij}$.
This is equivalent to exactly three ACFs of $\Pi_{ii}$ according to \eq~\eqr{eq:Q_ACF_2}.
This is convenient, because all dimensions should be sampled equally, and there are three dimensions, i.e., ACFs of $\Pi_{xx}$, $\Pi_{yy}$, and $\Pi_{zz}$ each contribute to the mean with a weight of $4/3$.
In all, there are ten equally contributing terms (four $Q_{ij}$ ACFs and six $\Pi_{ij}$ ACFs) in the sum, so a factor $1/10$ is needed to obtain the mean.
This gives, finally,
\begin{equation}
	\widetilde{\eta}(\nu) = \frac{\beta V}{10} \int_0^{\infty} \mathrm{d}t \,  
		e^{-i 2 \pi \nu t}
	\sum_{i, j}
	\left\langle
		\Pi_{ij}(t) \Pi_{ij}(0)
	\right\rangle \,.
	\label{eq:GreenKubo_full}
\end{equation}
The factor $1/10$ was first derived in Ref.~\refcite{1994_Daivis_JCP}.

The shear viscosity in the steady-state  limit $\eta_s$ may be found either by simply reading off $\widetilde{\eta}(\nu = 0)$, or equivalently, by taking the inverse Fourier transform of $\widetilde{\eta}(\nu)$ to obtain the viscosity $\eta(t)$ as a linear response function in the time domain,
\begin{equation}
	\eta(t) = 
	\frac{\beta V}{10}
	\sum_{i, j}
	\left\langle
		\Pi_{ij}(t) \Pi_{ij}(0)
	\right\rangle \,,
	\label{eq:visc_resp_fn}
\end{equation}
and taking the time integral from $t = 0$ to $\infty$,
\begin{equation}
	\eta_s = \frac{\beta V}{10} \int_0^{\infty} \mathrm{d}t \,  
	\sum_{i, j}
	\left\langle
		\Pi_{ij}(t) \Pi_{ij}(0)
	\right\rangle \,.
	\label{eq:visc_resp_fn_runn_int}
\end{equation}
In practice, the simulated ACFs of $\Pi_{ij}$ will be poorly behaved for long times due to insufficient data.
Instead, the running integral can be plotted, and should clearly plateau after correlations have vanished if there is sufficient data overall.
An average may then be taken over values in the plateau of the running integral.

\begin{figure}[!t]
	\centering
	\includegraphics{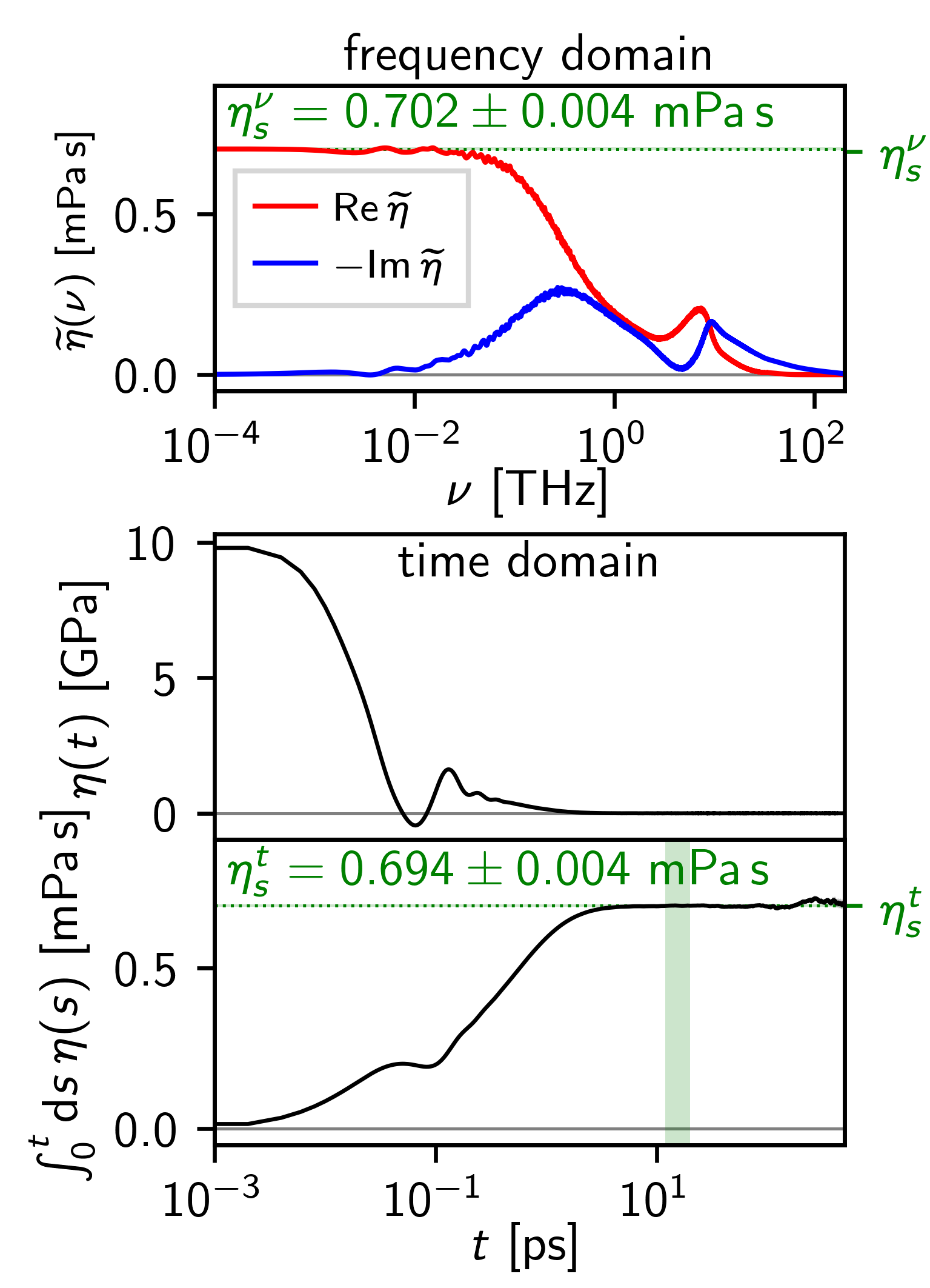}
	\caption{The viscosity spectrum calculated via \eq~\eqr{eq:GreenKubo_full} of an $NpT$ simulation of 4096 SPC/E water molecules in bulk with periodic boundary conditions at 300 K and 1 bar. 
	\textbf{Upper:} The real and imaginary parts of $\widetilde{\eta}(\nu)$ in the frequency domain.
	The zero-frequency value is shown as a dotted horizontal line and is denoted $\eta_s^\nu$.
	\textbf{Middle:} The Fourier transform of $\widetilde{\eta}(\nu)$, which is the response function $\eta(t)$. 
	\textbf{Lower:} The running integral of $\eta(t)$, which plateaus above about 10~ps.
	The plateau value, calculated in the green shaded region, is shown as a dotted horizontal line and is denoted $\eta_s^t$.
	\label{fig:GreenKuboVisc}}
\end{figure}


\begin{figure*}[!h]
	\centering
	\includegraphics{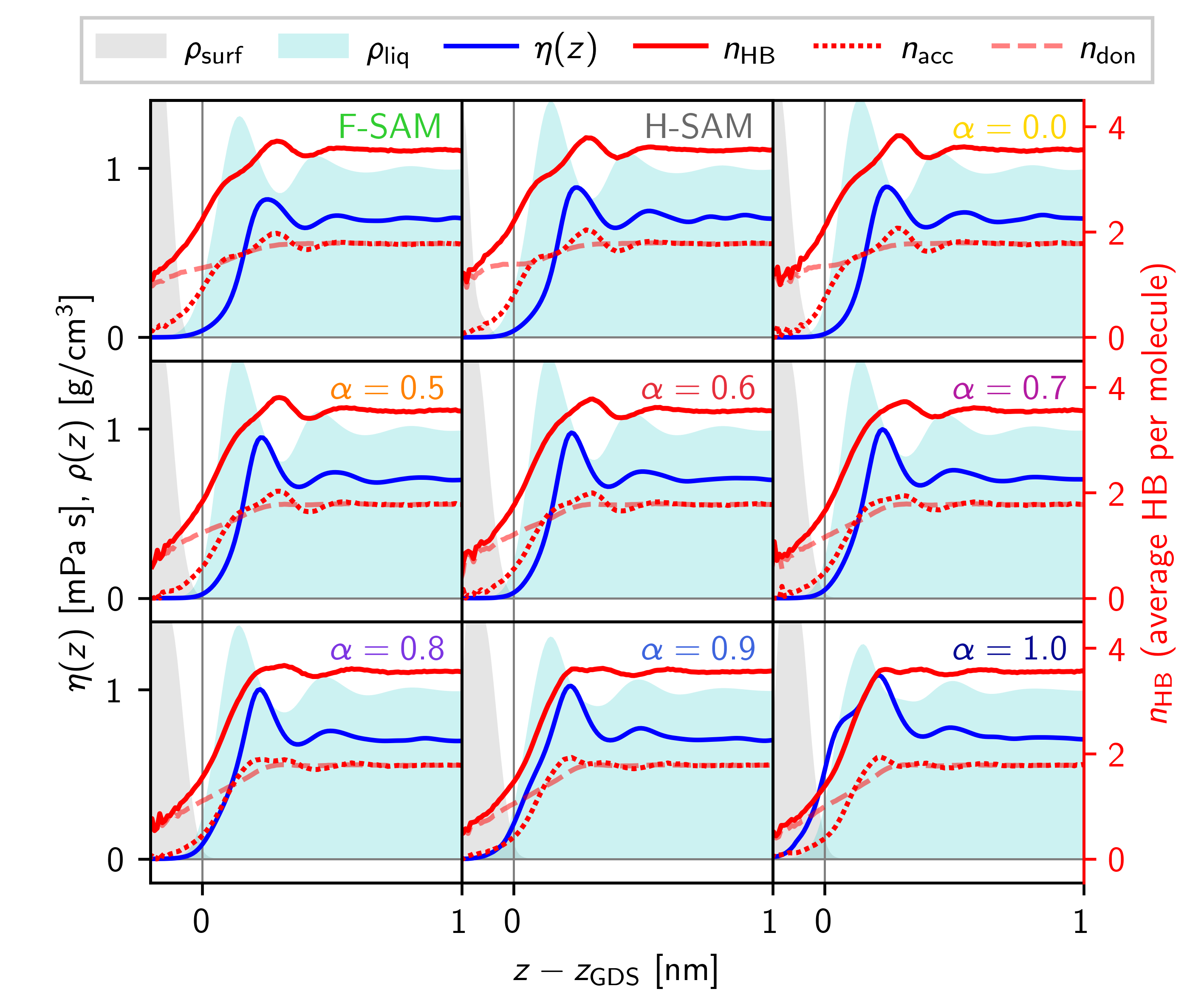}
	\includegraphics{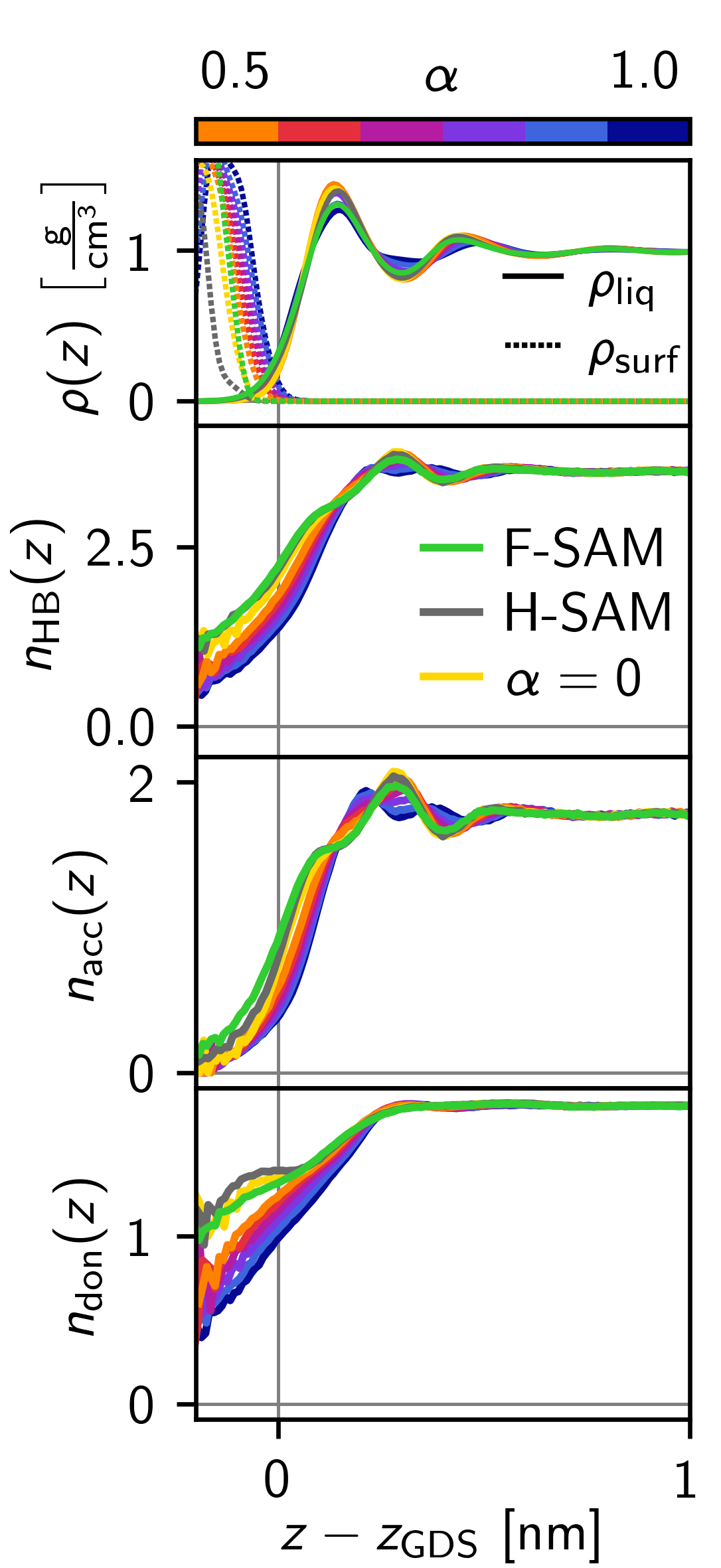}
	\caption{
	Water-water hydrogen bonds per molecule, $n_{\rm HB}(z)$, as a function of molecular center-of-mass $z$-position, as well as hydrogen bonds \emph{accepted}, $n_{\rm acc}(z)$, and \emph{donated} per molecule, $n_{\rm don}(z)$, for all systems studied in this work.
	\textbf{Left:} $n_{\rm HB}(z)$, $n_{\rm don}(z)$, and $n_{\rm acc}(z)$ compared to one another and to the viscosity profile $\visc(z)$ for each system separately.
	Mass density profiles of the surface and liquid, $\rho_{\rm surf}(z)$ and  $\rho_{\rm liq}(z)$, respectively, are shown as shaded areas as a positional reference.
	\textbf{Right:} $n_{\rm HB}(z)$, $n_{\rm don}(z)$, and $n_{\rm acc}(z)$ compared across systems.
	Here, $\rho_{\rm surf}(z)$ and  $\rho_{\rm liq}(z)$ are shown in the upper panel as a positional reference.
	\label{fig:HB_density}}
\end{figure*}

\begin{figure}[!h]
	\centering
	\includegraphics{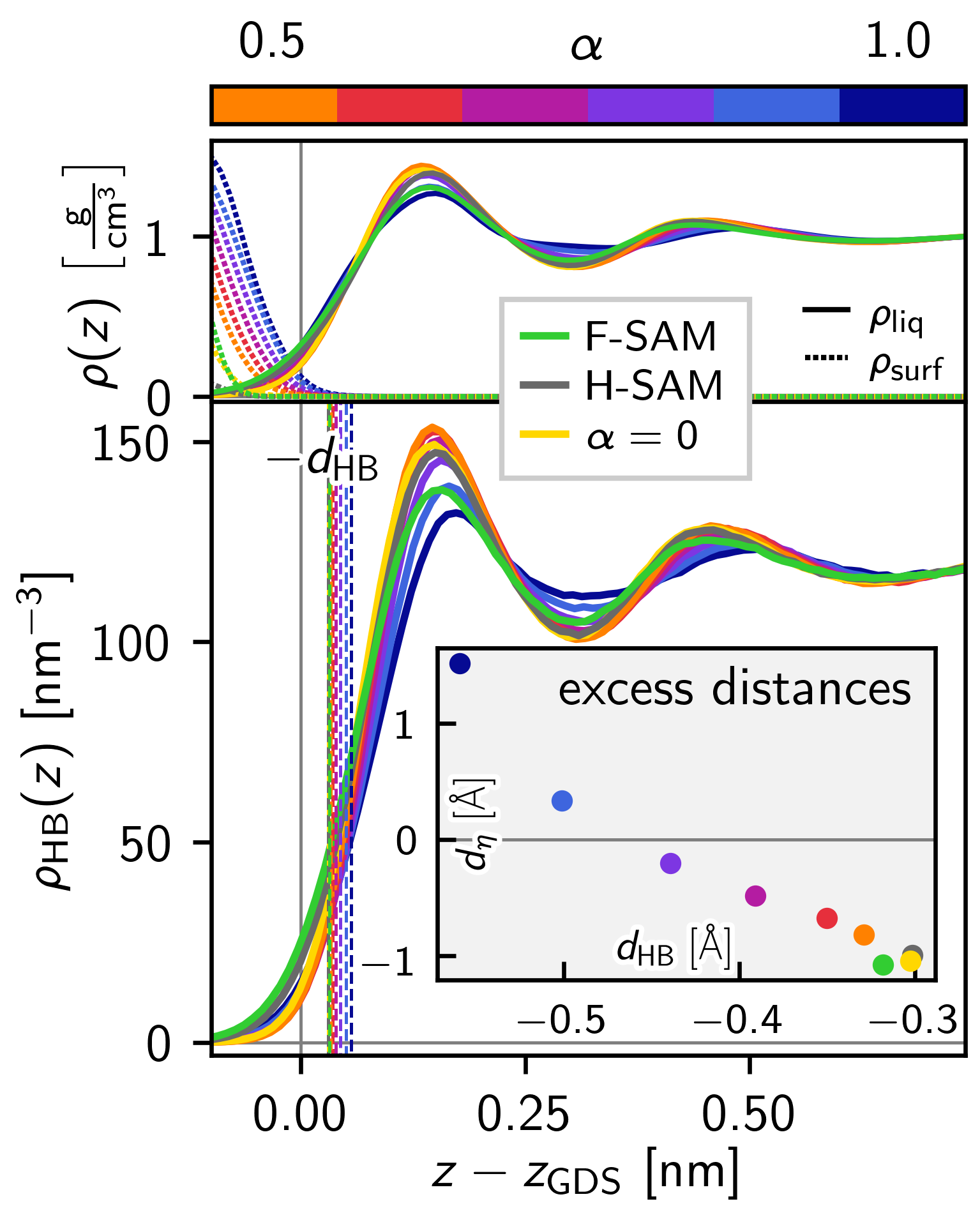}
	\caption{
	Water-water hydrogen bonds per unit volume, $\rho_{\rm HB}$, as a function of molecular center-of-mass $z$-position for all systems studied in this work. 
	Also shown as vertical dashed lines are the corresponding hydrogen-bond dividing surfaces. 
	Mass density profiles of the surface and liquid, $\rho_{\rm surf}(z)$ and  $\rho_{\rm liq}(z)$, respectively, are shown in the upper panel as a positional reference.
	The inset shows the interfacial excess distance of the viscosity, $d_{\visc}$, as a function of $d_{\rm HB}$.
	\label{fig:HB_density_exc}}
\end{figure}
\textbf{Results for SPC/E Water:} \fig~\ref{fig:GreenKuboVisc} shows the viscosity spectrum from an $NpT$ simulation of 4096 SPC/E water molecules with force-switching between 1.9 and 2.0~nm for Lennard--Jones forces at 300 K and 1 bar, calculated via \eq~\eqr{eq:GreenKubo_full}.
For pressure coupling, the exponential relaxation barostat was used (\texttt{C-rescale} in GROMACS) with time constant \texttt{tau-p} of 5 ps. 
The force fields used, including finite force cut-offs, are the same as for the driven-flow simulations in the main work.
The upper panel shows the real and imaginary parts of $\widetilde{\eta}(\nu)$. 
These are smoothed by convolution with a Gaussian kernel with $\sigma = 1.2$ GHz.
The resulting steady-state viscosity $\eta_s^\nu = \widetilde{\eta}(\nu = 0)$ is also shown as a horizontal dotted line.
The middle panel shows the response function $\eta(t)$.
The lower panel shows its running integral.
The steady state viscosity $\eta_s^t$, calculated as the average of the data in the plateau region (indicated by the vertical shaded green area), is shown as a horizontal dotted line. 
The two steady-state viscosities are formally identical, but due to numerical errors, they differ by about $1\%$. 
The value reported in the main text, $\eta_{\rm eq}=0.698$~mPa$\,$s, is the mean of $\eta_s^\nu $ and $\eta_s^t $.
For all data, error bars are shown as a shaded area above and below the curve, but these are not visible everywhere due to the errors being very small. 
The errors are calculated by breaking the simulation trajectory into six subtrajectories, and calculating the viscosity independently from each. 
The mean and error (standard deviation of the mean) can then be extracted for all quantities.

We also carried out the same procedure as shown in \fig~\ref{fig:GreenKuboVisc} on simulation data where a 0.9~nm Lennard--Jones cutoff was used, and found $\eta_{\rm eq}=0.698$~mPa$\,$s which agrees exactly with the value for force-switching between 1.9 and 2.0~nm.


\section{Hydrogen Bonding Near the Interface}
\label{sec:HB}
We measure the average number of water-water hydrogen bonds donated $n_{\rm don}$ and accepted $n_{\rm acc}$ per molecule as a function of the $z$-position of the molecular center of mass.
Following Ref.~\refcite{1996_Luzar_PRL}, we take two water molecules to be hydrogen bonded if their oxygen--oxygen distance $r_\mathrm{OO}$ is less than 3.5~{\AA}, and the (donor-hydrogen)--(donor-oxygen)--(acceptor-oxygen) angle, i.e., $\angle$HO$_D$O$_A$, is less than $30^\circ$.
\fig~\ref{fig:HB_density} plots $n_{\rm don}$, $n_{\rm acc}$, as well as the total $n_{\rm HB} = n_{\rm don} + n_{\rm acc}$.
The left part plots the data system-by-system, comparing it to the viscosity profiles.
For the hydrophobic surfaces, the viscosity profiles seem somewhat correlated with $n_{\rm acc}$ and/or $n_{\rm HB}$ across $z$, that is, the positions of the first and second peaks agree well.
This is unsurprising as hydrogen bonding is crucial for water cohesion, and can be expected to play a key role in the viscosity. 
This helps to explain the bulkward shift of the peak in the viscosity profile relative to the peak in the mass density in the first hydration layer.

The right part compares each profile type across all systems studied in this work, revealing an increase in interfacial water-water hydrogen bonds as surface hydrophobicity increases.
This is the opposite trend observed in the main text for the viscosity profiles $\visc(z)$.
This is studied more carefully in \fig~\ref{fig:HB_density_exc}, where the density of water-water hydrogen bonds per unit \emph{volume}, $\rho_{\rm HB}$, is plotted, again as a function of the $z$-position of the molecular center of mass. 
From $\rho_{\rm HB}(z)$, the interfacial excess distance of hydrogen bonds, $d_{\rm HB}$, may also be calculated in the same way as the interfacial excess of viscosity  $d_{\visc}$ is calculated in the main text. 
In \fig~\ref{fig:HB_density_exc}, $-d_{\rm HB}$ are plotted as vertical dashed lines.
The inset of \fig~\ref{fig:HB_density_exc} plots $d_{\visc}$ over $d_{\rm HB}$, and indeed, the two are negatively correlated across systems.
This implies that the excess surface viscosity near more hydrophilic surfaces must result from a different mechanism than an increase in water-water hydrogen bonding there.
We hypothesize the mechanism to be the conformational rigidity of water molecules hydrogen-bonded to the surface, which prevents other water molecules from easily flowing past.

\begin{figure}
	\centering
	\includegraphics{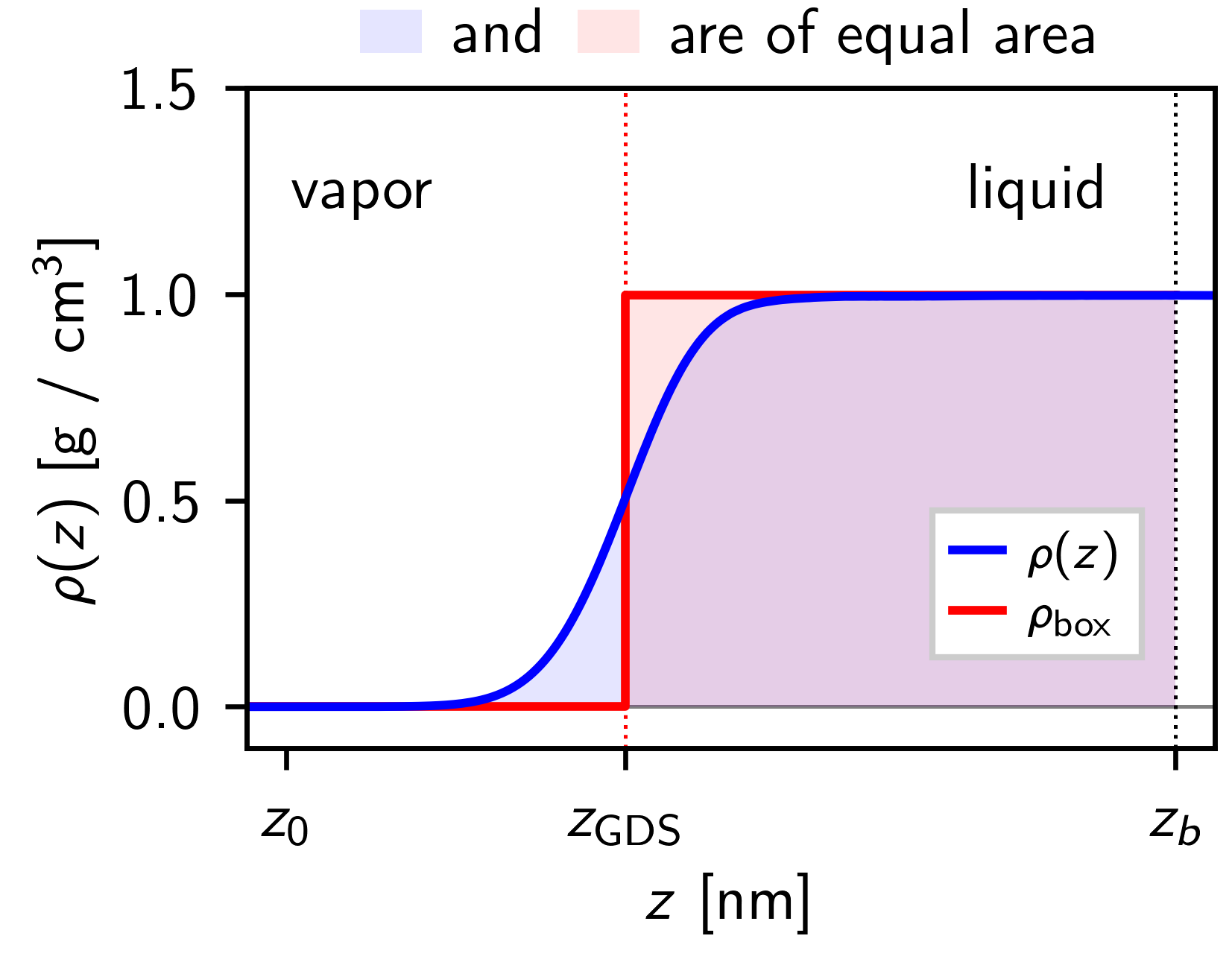}
	\caption{Plot of the density profile of a vapor-liquid water interface meant to illustrate the Gibbs dividing surface.
	A step-like box profile, with the step positioned at $z_{\rm GDS}$ is also shown.
	The Gibbs dividing surface is defined such that the red and blue shaded areas are equal, i.e., an integral over $\rho_{\rm liq}(z)$ from a position $z_0$ below the liquid, to a position $z_b$ in the bulk, will equal the integral over $\rho_{\rm box}(z)$ between the same boundaries.
	\label{fig:GDS_diagram}}
\end{figure} 

\section{Gibbs Dividing Surface}
\label{sec:GDS}

The Gibbs dividing surface defines the position of an interface separating two phases of matter, between which there may  in fact be a gradual change in the density and/or composition over some finite distance (usually on the order of Angstroms or nanometers).
We consider the case of an interface separating a vapor and liquid phase as pictured in \fig~\ref{fig:GDS_diagram}.
The position of the Gibbs dividing surface, $z_{\mathrm{GDS}}$, is taken as the position where the surface excess vanishes, i.e.,
\begin{equation}
	z_{\mathrm{GDS}} 
	= 
	z_0 
	+ \int_{z_0}^{z_b} \mathrm{d}z \,
		\frac{\rho(z_b) - \rho(z)}{\rho(z_b) - \rho(z_0)} \,,
	\label{eq:GDS}
\end{equation}
where $\rho(z)$ is the density profile along $z$, and $z_0$ and $z_b$ are positions in the vapor and liquid bulk phases, respectively, well away from the interface \cite{2012_Bonthius_Langmuir}.


\section{Depletion Length}
\label{sec:DepLen}

There is typically a zone of low density at surface--liquid interfaces called the depletion layer, the width of which is referred to as the depletion length $\delta$.
The depletion length can be quantified by an integral over the relative mass density deficit. 
We are interested in the depletion layer specifically as a zone where there is a deficit of occupation by atoms, i.e., where less \emph{space} is taken up on average than in the bulk of the solid or liquid.
Thus, we define the depletion length with respect to the packing density $\packdens$, instead of the usual mass density $\rho$. 
Here, the packing density $\packdens$ is defined as the unitless ratio of the total van der Waals volume of atoms within a region $\Omega$ to the total volume $V_\Omega$ of that region, i.e.,
\begin{equation}
	\packdens(\Omega) 
	= 
	\frac{1}{V_\Omega} 
	\sum_{i \in \Omega} \frac{4}{3} \pi \left( r_i^{\mathrm{vdW}} \right)^3 \,,
\label{eq:packdens_denf}
\end{equation}
where $r_i^{\mathrm{vdW}}$ is the van der Waals radius of the $i^\text{th}$ atom in $\Omega$.
The van der Waals radii used in this work are experimental values collected in Ref.~\refcite{1996_Rowland_JPC}, and are reproduced in Table \ref{tab:vdW_radii_of_SAM_atoms}.
\begin{table}[!h]
    \centering
    \begin{tabular}{p{15mm} p{10mm} p{10mm} p{10mm} p{10mm}}
    \hline
     atom &  C &  H & O & F  \\
    \hline 
	\rule{0pt}{2ex} $r^{\mathrm{vdW}}$ [{\AA}] & 1.77 &  1.1 & 1.58 & 1.46  \\
    \hline
    \end{tabular}
    \caption{Experimental van der Waals radii as published in Ref.~\refcite{1996_Rowland_JPC}.
    \label{tab:vdW_radii_of_SAM_atoms}}
\end{table}

\begin{figure}
	\centering
	\includegraphics{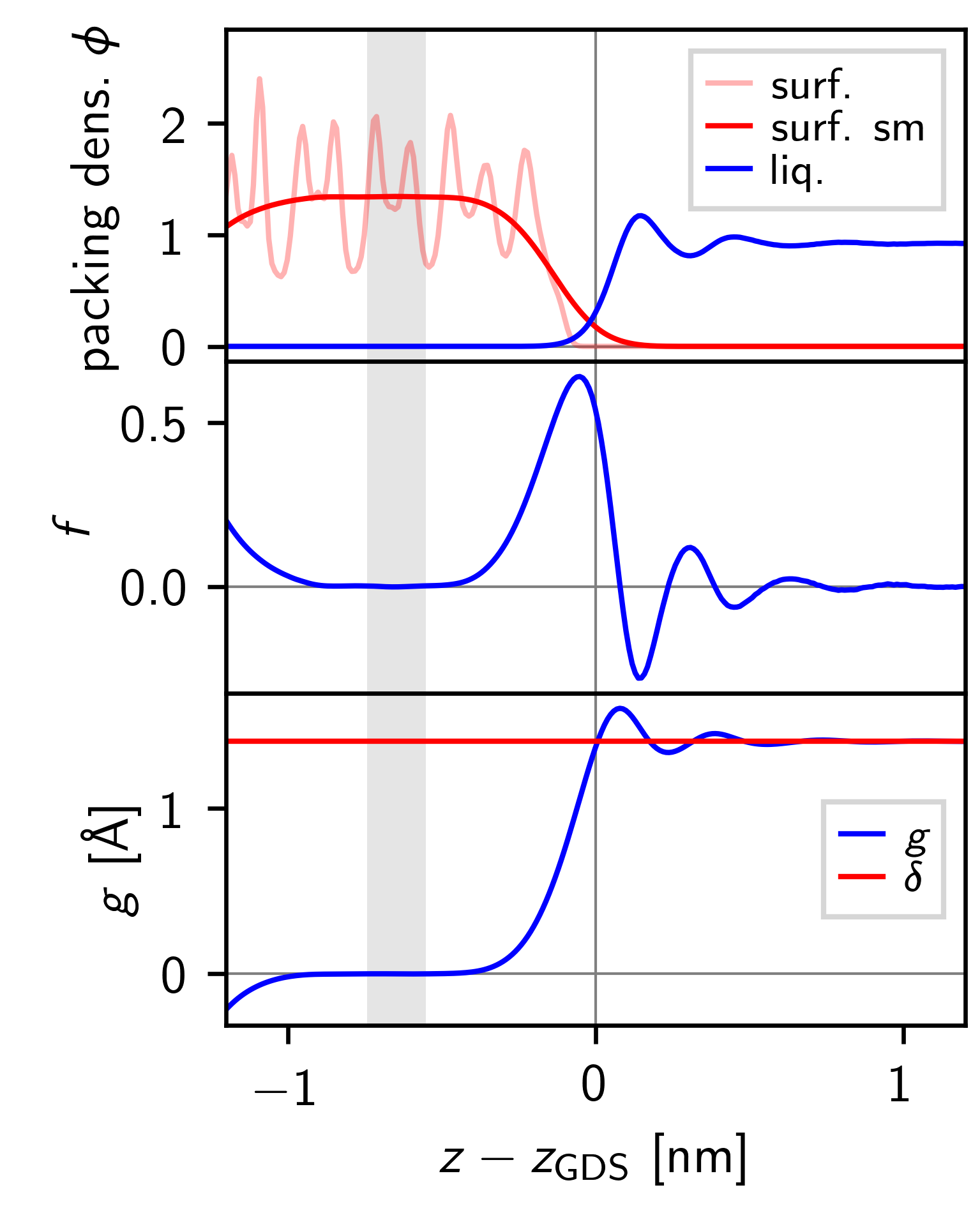}
	\caption{Plots illustrating the extraction of the depletion length for the F-SAM/water system. 
	The upper panel shows packing density profiles for the SAM and water,  $\packdens_s$ and  $\packdens_l$, respectively.
	The density profile $\packdens_s$ of the SAM is smoothed with a Gaussian convolution, which preserves the integral, in order to find the bulk density, which is calculated as the mean in the gray shaded region.
	The middle and lower panels show $f$ and $g$ (see \eqs~\eqr{eq:depletion_length} and \eqr{eq:depletion_g}, respectively).
	The red horizontal line in the lower panel is the resulting depletion length $\delta$.
	\label{fig:depletion_extraction}}
\end{figure}

For homogeneous phases, such as pure liquids, the mass density and packing density give identical results, but for non-homogeneous phases, such as a SAM with terminal OH-groups, they give different results. 
Thus, the depletion length is defined as
\begin{align}
	\delta &= \int_{z_s}^{z_l} \mathrm{d}z \, f(z)  \notag \\
	 &= \int_{z_s}^{z_l} \mathrm{d}z \,
	\left( 
		1 - \frac{\packdens_s(z)}{\packdens_s^b} - \frac{\packdens_l(z)}{\packdens_l^b}	
	\right) \,,
	\label{eq:depletion_length}
\end{align}
where $\packdens_s(z)$ and $\packdens_l(z)$ are the surface and liquid packing density profiles, $\packdens_s^b$ and $\packdens_l^b$ are their respective bulk values, and $z_s$ and $z_l$ are $z$-positions well within the bulk of the surface and liquid respectively
\cite{2004_Mamatkulov_Langmuir, 2007_Maccarini_Langmuir, 2007_Janecek_Langmuir, 2008_Sedlmeier_Biointerphases, 2009_Sendner_Langmuir}. 
The integrand $f(z)$ is the density deficit and its running integral we define as 
\begin{equation}
	g(z) = \int_{z_s}^z \mathrm{d}z^\prime \, f(z^\prime) \,.
	\label{eq:depletion_g}
\end{equation} 
The upper panel of \fig~\ref{fig:depletion_extraction} shows plots of packing densities of liquid water and an F-SAM. 
The density of the SAM is smoothed by convolution with a Gaussian in order to facilitate determination of $\packdens_s^b$. 
Note that the Gaussian convolution preserves the integral of the density. 
The middle and lower panels show the corresponding ${f(z)}$ and ${g(z)}$. The mean value for the SAM density $\rho_s^b$ is taken over the shaded region in the plot. 
The depletion length $\delta$  may be calculated from the difference in the bulk values of ${g(z)}$ in the solid and liquid, as shown in the figure.

\section{The Friction--Wettability Relationship}

The motion of an interfacial liquid molecule tangential to a surface can be treated as a series of barrier crossings over a corrugated potential landscape. Take the potential landscape $U(x)$ to be a periodic potential with a minimum at $x=0$, amplitude $U_0$, and $2L$-periodicity. A particle in a well has to cross a barrier of height $U_0$ to move one well to the left or right.
From transition state theory, the mean barrier crossing time $\tau$ over a barrier of height $U_0$ is 
\begin{equation}
	\tau = \tau_0 e^{\beta U_0}\,,
	\label{eq:bc_v_of_tau}
\end{equation}
where $\beta = (k_B T)^{-1}$ is the inverse thermal energy and $\tau_0$ is a constant \cite{1940_Kramers_P, 2022_Bruenig_PRE}.
Consider the case of a small, constant force $F$ applied to the particle ($L F \ll U_0$, $L F \ll \beta^{-1}$). This changes the barrier heights on the left and right of the particle to $U_l \approx U_0 + LF$ and $U_r \approx U_0 - LF$. Letting the average rate of barrier crossings to the left and right be given by $\tau_l$ and $\tau_r$, the average velocity of the particle is
\begin{align}
	v &= 2 L \left( \frac{1}{\tau_r} - \frac{1}{\tau_l} \right) \notag \\
	&= \frac{2L}{\tau_0} \left( e^{-\beta U_r} -  e^{-\beta U_l} \right) \notag \\
	&\approx \frac{2L}{\tau_0} 
	\left( 
		e^{-\beta (U_0 - LF) } -  e^{-\beta (U_0 + LF)} 
	\right) 
	\notag \\
	&= \frac{2L}{\tau_0} e^{-\beta U_0 } 
	\left( 
		e^{\beta LF } -  e^{-\beta LF} 
	\right) 
	\notag \\
	&= \frac{4L}{\tau_0} e^{-\beta U_0 }  \sinh \left(\beta LF \right) 
	\notag \\
	&\approx \frac{4L}{\tau_0} e^{-\beta U_0 } \beta LF
	\,,
	\label{eq:bc_v_derivation}
\end{align}
where in the final step it is used that $\sinh x \approx x$ for small $x$ up to third order. Let there be $n$ interfacial particles per unit area, then the stress driving the velocity $v$ is $-F_f = nF$.
Substituting and rearranging gives
\begin{equation}
	F_f = -\frac{n \tau_0 e^{\beta U_0}}{4 \beta L^2}  v \,,
\end{equation}
which gives for the Navier friction coefficient
\begin{equation}
	\fks = \frac{N \tau_0 e^{\beta U_0} }{4 \beta L^2 } \propto 			e^{\beta U_0}\,.
	\label{eq:bc_lambda}
\end{equation}
Here it is assumed that $n$ is constant in $U_0$, which we justify by saying that all liquid molecules directly adjacent to the surface see the corrugated potential landscape, and this number is relatively constant over the different systems studied in this work.
The potential landscape $U(x)$ seen by the particle is in fact the sum of all surface--liquid interactions along its tangential trajectory. Therefore, it is should hold that $U_0 \propto \varepsilon_\mathrm{int}$, where $\varepsilon_\mathrm{int}$ is the areal surface--liquid interaction energy. 
Young's equation relates the areal work of adhesion $W$ to the wetting coefficient $k=\cos \theta$ (where $\theta$ is the contact angle) and liquid--vapor interfacial tension $\gamma$,
\begin{equation}
	W = \gamma \left( k + 1 \right) \,.
	\label{eq:bc_youngs}
\end{equation}
In the approximation $W \propto \varepsilon_\mathrm{int}$, \eqs~\eqr{eq:bc_lambda} and \eqr{eq:bc_youngs} give a scaling relation between the steady-state Navier friction coefficient $\fks$ and the contact angle
\begin{equation}
	\fks \propto e^{A \left(k + 1\right)} \,.
	\label{eq:bc_lambda_theta_scaling_rel}
\end{equation}
In Ref.~\refcite{2012_Erbas_JACS}, scaling relationships are derived starting from the Fokker-Planck equation and assuming the corrugated potential landscape
\begin{equation}
	U(x) = -U_0 \frac{1 - \cos\left( \frac{\pi x }{L} \right)}{2} \,.
\end{equation}
The result for the friction coefficient is 
\begin{equation}
	\fks \propto 
	\int_{0}^{2L} \mathrm{d}x \, e^{\beta U\left( \frac{x}{2L} \right)}
	\int_{0}^{2L} \mathrm{d}x \, e^{-\beta U\left( \frac{x}{2L} \right)} \,.
\end{equation}
In the high-barrier limit ($U_0 \to \infty$) this gives 
\begin{equation}
	\fks \propto \frac{e^{\beta U_0}}{U_0}\,.
	\label{eq:bc_lambda_theta_scaling_rel_EXPOX}
\end{equation}
In the low-barrier  limit ($U_0 \to 0$) it gives, to third order,
\begin{equation}
	\fks \propto 1 + \frac{\left( \beta U_0 \right)^2}{16}\,.
	\label{eq:bc_lambda_theta_scaling_rel_QUAD}
\end{equation}
This quadratic relationship has also been explored in other publications \cite{2009_Huang_PRL, 2009_Sendner_Langmuir}.

\begin{figure}[!t]
	\centering
	\includegraphics{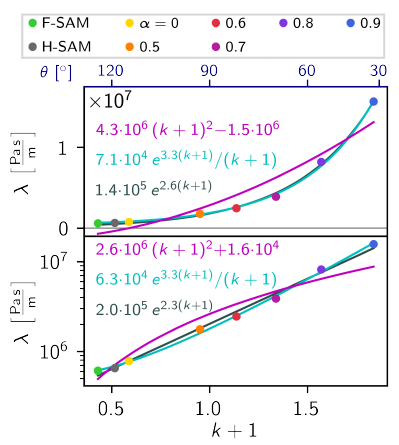}
	\caption{A comparison of exponential (\eqs~\eqr{eq:bc_lambda_theta_scaling_rel} and \eqr{eq:bc_lambda_theta_scaling_rel_EXPOX}) and quadratic (\eq~\eqr{eq:bc_lambda_theta_scaling_rel_QUAD}) fits of the Navier friction coefficient $\fks$ over $k+1$, where $k=\cos \theta$ is the wetting coefficient, for all systems studied in this work, aside from the ($\alpha$=1)-SAM/water system. 
	While the data plotted in the two panels are the same, the scale of the $y$-axis is linear for the upper panel and logarithmic for the lower panel, and the fit in the lower panel was performed in logarithmic space.
	\label{fig:fric_wett_exp_vs_quad_SI}}
\end{figure}
\fig~\ref{fig:fric_wett_exp_vs_quad_SI} compares fits of \eqs~\eqr{eq:bc_lambda_theta_scaling_rel}, \eqr{eq:bc_lambda_theta_scaling_rel_EXPOX} and \eqr{eq:bc_lambda_theta_scaling_rel_QUAD}  to the friction-wetting data for all systems studied in this work. 
In the upper panel of \fig~\ref{fig:fric_wett_exp_vs_quad_SI}, which has a linear $y$-axis, the data are fit directly with the fitting functions.
In the lower panel, which has a logarithmic $y$-axis, the fit parameters are obtained in logarithmic space, i.e., by fitting the logarithm of the $y$-data with the logarithm of the respective fitting function.
The resulting fitted functions are printed directly in the corresponding panels.
Table~\ref{tab:fric_wett_fit_residuals} shows the sum of squared residuals of the fits.
The exponential fits (\eqs~\eqr{eq:bc_lambda_theta_scaling_rel} and \eqr{eq:bc_lambda_theta_scaling_rel_EXPOX}) seem to capture the behavior much more accurately than the quadratic fit, as is also apparent from Table~\ref{tab:fric_wett_fit_residuals}.
This is perhaps to be expected as the low-barrier limit is a poor approximation for hydrophilic surfaces like the ($\alpha$=0.8)- and ($\alpha$=0.9)-SAMs.
The two exponential fits, (\eqs~\eqr{eq:bc_lambda_theta_scaling_rel} and \eqr{eq:bc_lambda_theta_scaling_rel_EXPOX}) seem to fit the data with a roughly similar degree of accuracy, with Table~\ref{tab:fric_wett_fit_residuals} indicating that the fit of \eq~\eqr{eq:bc_lambda_theta_scaling_rel_EXPOX} is marginally more accurate.
As the fits are of similar quality,  we reproduce the fit of \eq~\eqr{eq:bc_lambda_theta_scaling_rel} in \fig~{\mainFigAnalysis} in the main text, as the simple exponential function better facilitates comparison with the depletion length and viscosity excess distance.
\begin{table}[!t]
    \centering
    \begin{tabular}{p{37mm} p{16mm} p{16mm}}
    \hline
     & fit of $y$ \newline (upper) & fit of $\log y$ \newline (lower)   \\
    \hline 
	\eq~\eqr{eq:bc_lambda_theta_scaling_rel}: $A e^{B(k+1)}$  & 7.134$\times$10$^{10}$ & 1.106$\times$10$^{-2}$   \\ 
	\eq~\eqr{eq:bc_lambda_theta_scaling_rel_EXPOX}: $A e^{B(k+1)}/(k+1)$  & 6.902$\times$10$^{10}$ &  3.724$\times$10$^{-3}$ \\
	\eq~\eqr{eq:bc_lambda_theta_scaling_rel_QUAD}: $A (k+1)^2 + B$ & 3.248$\times$10$^{12}$ & 1.144$\times$10$^{-1}$\\ 
    \hline
    \end{tabular}
    \caption{The sum of squared residuals for the fits shown in \fig~\ref{fig:fric_wett_exp_vs_quad_SI}. 
    For the right column, the residuals are calculated in logarithmic space.
    \label{tab:fric_wett_fit_residuals}}
\end{table}

\begingroup
\footnotesize
\bibliography{bibliography.bib}
\endgroup